\documentstyle[12pt,aasms4]{article}
\begin{document}

\title{The Dynamics of the cD Clusters Abell 119 and Abell 133}

\author{M.J. Way\altaffilmark{1}, H. Quintana\altaffilmark{2,3} and
L. Infante}

\affil{Department of Astronomy and Astrophysics,
P. Universidad Catolica de Chile,\\
Casilla 104, Santiago 22 , Chile\\ I: mway@newton.umsl.edu}

\altaffiltext{1}{present address: Department of Physics and Astronomy,
University of Missouri-St.Louis, 8001 Natural Bridge Rd., St.Louis, 
MO 63121-4499}
\altaffiltext{2}{Visiting Astronomer, Las Campanas Observatory
of The Carnegie Institute of Washington} 
\altaffiltext{3}{1995 Presidential Chair in Science} 

\begin{abstract}

	A dynamical analysis of the structure of the clusters of galaxies Abell
119 and Abell 133 is presented, using new redshift data combined with existing
data from the literature. We compare our results with those from the X-ray data
for these clusters, and with those from radio data for the central cD galaxy in
each cluster. A comparison of the mass estimate based on X-ray data and that
obtained here after subgroups are eliminated shows them to be comparable.
After the elimination of subgroups, 125 galaxy members in Abell 119 and
120 in Abell 133 give dispersions of 472 km s$^{-1}$ and 735 km s$^{-1}$
respectively. However, our dynamical analysis of the optical data shows little
substructure in the velocity field of Abell 133, conflicting with what is seen
in the Rosat X-ray map.  Abell 119 seems to have multiple structures
along the line of sight.  We derive virial mass estimates of 3.05 X 10$^{14}$
M$_{\sun}$ for Abell 119 and 7.79 X 10$^{14}$ M$_{\sun}$ for Abell 133 within
1.5 h$^{-1}$Mpc, which agree well with the X-ray-derived masses within errors.
\end{abstract}


\section{Introduction}

Regular, rich clusters of galaxies are the largest objects in the Universe
likely to be bound and possibly relaxed. Their differing shapes, concentrations
and populations are customarily interpreted as  representing different stages
in the gravitational evolution of the matter of the cluster.
The most regular clusters
usually have galaxies of type cD  at their centers. The view that such clusters
have reached stationary equilibrium is generally accepted.  However, recent
work has shown the presence of significant substructure in
these clusters, challenging
the virialization and relaxation hypothesis. To search for
substructure one need not only look at the distribution of
galaxies, but to analyze the gas profiles and inhomogeneities as revealed by
the X-ray data and the dynamical information provided by the velocity field of
the galaxies throughout the cluster.

The nearby cD clusters Abell 119 and Abell 133 are among the brightest
sources in X-rays.
As such, they were promising sources to be studied in depth in the optical
and X-ray bands. Together with other clusters accessible to our southern
programs, they were chosen to carry out a deeper velocity survey using
multi-fiber spectrographs. Here we report velocities taken with the Las
Campanas DuPont 100'' telescope. 

	In Section II we discuss how the data were obtained. We then explain
the data reduction in section III, and in IV we detail how 
velocity data from the literature compares with this paper's data, and how it
was integrated to our data for analysis. The dynamical analysis is presented in
section V, broken up into several subsections for Abell 119 and 133. Finally,
section VI contains a discussion and the implications of our analysis.

\section{Observations}

The aim of our program was to obtain velocities for as many galaxies as
possible over a reasonably  wide field, within the observational allocation of
time. We chose to survey areas of 1.5$\arcdeg$x1.5$\arcdeg$ centered
on the cD
galaxies, the field covered by 100'' DuPont telescope fiber spectrograph.
Lacking previous  photometric data to select a magnitude limited sample, we
measured the positions of galaxies clearly  identified on the glass copies of
the Palomar Sky Survey at ESO, Garching. The plates were searched by eye on
the Optronics machine monitor following strips in declination. The final
astrometric lists contained  some 400 galaxies in each cluster and a number of
stars used for guiding the fiber arrays. The  positions were
determined from astrometric solutions based on 25 SAO or Perth reference stars,
using  standard programs at ESO. Their relative accuracy is 0.3 arcsec rms.
However, the external accuracy  should be of order 2 arcsec. 
See figures 1a and 1b for plots of the x-y positions. Notice in
figure 1b that the bottom of the plot is blank because the entire field
of Abell 133 is close to the edge of the scanned Palomar plate. We could not use
a second plate to obtain the rest of the velocities because the external
positional error between plates was too high.

	We used Shectman's fiber spectrograph (\cite{SH85}) 
mounted on the 100'' telescope on the
nights of 22-25 October 1990. The multi-fiber system consists of a plug  plate
at the focal plane to which 65 fibers are attached and run to a Boller and
Chivens spectrograph coupled to a 2DFrutti detector (2DF). A 600 line mm$^{-1}$
grating blazed at 5000{\AA} was set at an angle $9^{\circ}$  40', giving a
wavelength coverage from $\sim$3800-6800{\AA}. Normally, 50-55 fibers are used
for  objects. Ten sky fibers are set aside, spaced at intervals of one
every 6 fibers along the spectrograph entrance, and positioned in a 
random pattern in the plug plate. The resulting 2DF  image has a 1520 x
1024 pixel area, with a dispersion of $\sim$2.6{\AA} pixel$^{-1}$ and
a final resolution of $\sim$10{\AA}.
The fiber images are $\sim$8 pixels wide, and separated by $\sim$12 pixels from
center to center. 

Exposure times were adjusted to be between 80 and 120 minutes, depending on the
brightness of the selected galaxies for each exposure. The 2DF detector is
a photon counting system where one can view the current exposure at any stage.
In this way one can obtain the optimum exposure time for a field.  A faintness
limit of $\sim$17.5 magnitudes in R was reached.  150 spectra in each of
Abell 133 and Abell 119 were obtained in this run.  Quartz lamp exposures were
used to correct for pixel to pixel variations of the detector. To
properly illuminate the whole detector surface the grating angle was changed
to several values on these exposures. As well, helium-neon  comparison lamp
exposures were taken off the wind-screen for wavelength calibration before and
after each exposure. The 2DFrutti detector has a small dark current and no
corrections were made for that effect.

\section{Reductions}

Velocity determinations were carried out using a cross-correlation technique
{\it and} by identifying and fitting by eye line profiles. All reductions
were performed inside of the IRAF(\cite{T93}) environment.
For a complete discussion of the reductions see \cite{QRW96}(hereafter QRW96).
That described below
is a summary of the reductions. Due to the nature of the fiber+2DFrutti
system typical S shaped distortions are inherent in this instrument.  A sixth
order spline3 curve was used to trace the S shaped spectra.  The IRAF HYDRA
package was used to extract the spectra, correct pixel to pixel variations
via the dome flat, use a Fiber Transmission Table for appropriate sky
subtraction and put the spectra on a
linear in wavelength scale. The wavelength solutions for 20-30 points using a
5, 6, or 7 order Chebyshev typically yielded residual values less than 0.4 rms
{\AA}, where 1 pix $\sim$2.6{\AA}.  The ten sky spectra from each exposure
were combined via a median filter and subtracted from each of the object
spectra. 

Two different methods were used to measure the redshift of the objects.
For normal early type spectra the RVSAO (\cite{K91}) cross-correlation algorithm
supported inside IRAF was used. The algorithm used in RVSAO is described
in \cite{TD79} (hereafter TD79). A reliability factor was generated by
RVSAO called the R value (see TD79 for details).  Normally a low R value
(R$\leq{4}$) indicated a need to look at the spectra and try line by line
Gaussian fitting (the second method). To utilize RVSAO {\it template}
spectra with high signal to noise and well determined radial velocities
were needed. Two of the four templates used in this paper were galaxy
spectra taken with the fiber instrument, NGC 1407
and NGC 1426, another galaxy (NGC1700) was from the previous detector on the
2.5 meter at Las Campanas (Shectograph), and one was a synthetic
template. The synthetic template was constructed from the excellent library
of stellar spectra of \cite{JHC84}. We used ratios of
stellar light for the E0 galaxy NGC1374 from the synthesis
studies of \cite{PIC85}. In the end it was found that the
template which gave the lowest error value out of the four radial velocity
cross-correlation templates mentioned above proved to have more consistent
results.  For non-early type spectra (i.e emission lines, E+A, etc.)
a line by line Gaussian fit was used and the resulting
velocities from each line fit were averaged. 

Tables 1 (Abell 119) and 2 (Abell 133) give the velocity results,
where column 1 shows the identification number,
columns 2 and 3 give 1950.0 epoch positions, and columns 4, 5, 6, 7 and 8
give the individual velocity values, their errors (1 $\sigma$),
the corresponding
TD79 R values or number of measured lines (mostly emission) if
the R number was too low for a proper velocity determination, references,
and the identification number from each reference.  
Columns 9 and 10 give the final adopted velocities, if averaged, and their
respective one $\sigma$ errors.

Multiple measurements and integrated velocity values from the literature were
combined and are described in the next section. Where large discrepancies
exist with values quoted in the literature, they have been eliminated
as indicated in the
comments. The individual velocities retained have been averaged and weighted
by the corresponding quoted errors (combined in quadrature) for an estimate
of the final error.

For a number of galaxies 2 spectra were measured. This allows a check on
internal consistency. These velocities are listed for the 
corresponding galaxies in Tables 1 and 2. In
Abell 133 there are six galaxies with two measurements which differ by values
from 6 to 75 km s$^{-1}$, just within the errors, showing good internal
consistency.  In Abell 119 five galaxies have 2 fiber spectra with differences
ranging from 7 to 117 km s$^{-1}$. The only galaxy not falling within the
quadratically added errors was number 25370 where the difference was 117 km
s$^{-1}$ which can be attributed to the low signal to noise in one of
the spectra obtained.

\section{Comparison}

A zero point shift for references with several
velocities in common can be used to bring the data to a common
system. This has been applied before averaging weighted by the
respective errors (see QRW96 for details). 
The first systematic determination of velocities in Abell 119 was done by
\cite{MQ81} (hereafter MQ81), who published 23 velocities, with typical errors
100-250 km s$^{-1}$. For 19 galaxies a systematic shift was found
between this paper (fibr) and MQ81
(figure 2a) of $\sim$-45 $\pm$ 217 km s$^{-1}$ rms after applying a $\sigma$
clipping routine that rids one of those MQ81 velocities quoted with
errors larger than 150 km s$^{-1}$ (with the exception of one).
As noted in MQ81, errors larger than 150 km s$^{-1}$ denote very uncertain
velocities, a fact borne out by the new measurements. It is also noted that the
position of galaxy 10 in MQ81 is wrong (due to a re-numbering of galaxies
in that paper). \cite{FAB93} (hereafter FAB93) give values for 60 velocities. 
Of the FAB93 velocities 48 are in common with this paper's fiber data.
In comparison
with FAB93 (see figure 2b) there is a systematic shift of $\sim$-28 km s$^{-1}$
with an RMS of $\sim$76 km s$^{-1}$. For completeness, in Table 1 12 galaxies
are included with velocities measured solely by FAB93.  Several velocity
measurements in Abell 119 are added from the papers of \cite{ZAB93}
(2 measurements), \cite{RC3} (4), \cite{HDTL} (1), \cite{S78} (1), and
\cite{KH} (1).  A comparison with the two velocities in common with
\cite{ZAB93} show a systematic shift of 5 $\sim$-45 $\pm$ 86 km s$^{-1}$ rms.
In Abell 133 overlapping velocities were found in the papers of
\cite{MK} (3), and \cite{RC3} (5).  In \cite{MK} a shift of 
-134 km s$^{-1}$ $\pm$ 64 rms was found.  Using the 9 velocities
in common with Abell 133 and Abell 119 from \cite{RC3} a systematic
shift of -34 km s$^{-1}$ $\pm$ 81 rms was found.
All of the shifts above suggest that it is this paper's data which is shifted
by roughly 40 km s$^{-1}$ with respect to the literature.

\section{Dynamical Analysis}


\subsection{Abell 119}

\subsubsection{1D tests (velocity space)}

	Information from the 1D velocity distribution
was obtained with ROSTAT, a program for robust estimation of
velocity distributions based on the work of \cite{BFG90} and kindly distributed
by Tim Beers. ROSTAT was first used to calculate the robust estimators of
location C$_{BI}$ (average) and scale S$_{BI}$ (dispersion) with 10000
bootstraps and with 90\% confidence intervals (see Table 3).
As a first order attempt to remove outlier galaxies in the velocity
distribution a standard 3 $\sigma$ (S$_{BI}$) clipping of \cite{YV77} was used.
Consequently, all galaxies in Abell 119
closer than 11033 and further than 15609 km sec$^{-1}$ were eliminated.
This 3$\sigma$ clipped data is taken as defining {\it the cluster}.
Then {\it the cluster} data is taken through a second ROSTAT run
using those points within the velocity ranges above to utilize the shape
estimators of skewness, kurtosis, tail and asymmetry. Their values
are listed in Table 3. For a thorough discussion of these estimators
see \cite{BB93}.

	Of the 4 estimators only the robust estimator of asymmetry shows no
strong deviation from a Gaussian distribution, but note that the robust
estimators, asymmetry and tail, are
normally much more conservative and will give fewer false positives
for non-Gaussian distributions than the classical skewness and kurtosis.
The high positive kurtosis
(leptokurtic) value leads one to believe the tails are heavier than expected
for a Gaussian. The large positive value of skewness implies
a lack of values below C$_{BI}$ or that values on the positive side
of C$_{BI}$ are more enhanced than a standard Gaussian distribution.
The tail index points to a double exponential shape rather than Gaussian.
Given these indicators of non-Gaussian behavior the KMM objective
partitioning algorithm was used to search for multiple Gaussians
in the velocity field. See \cite{A94} for a detailed discussion.
Briefly, KMM fits a user-specified number of Gaussians to the velocity data
and estimates the improvement of the multiple Gaussian fit versus
a single Gaussian. The user first inputs an estimate of the positions
of the multiple Gaussians. Using the velocity histogram (figure 3a) as a 
starting point two, three, four, and five Gaussians were fit to the data.
For the three Gaussian fit, Gaussians were first estimated at
11900, 13200, and 14650 km sec$^{-1}$ from the velocity histogram.
The program returned values of 11837,
13253 and 14718 km sec$^{-1}$ with a rejection of the single Gaussian
model at a confidence level of 97.9\%. This three Gaussian model implies that
two smaller groups are projected or in-falling toward the main cluster body,
one from the front and one from the back. In fact
FAB93 detected the foreground group as having velocities less than
12000 km sec$^{-1}$, but miss the background group. In figure 4 the three
groups are over plotted. {\it Group 1} is denoted by $\Box$ (11 members),
{\it group 2} by $\bullet$ (125 members) and {\it group 3} by + (17 members).
Further attempts to subdivide {\it group 2} using KMM were unsuccessful.
Running ROSTAT on the 3 groups provides one with further information.
{\it Group 1} has a large negative skewness implying a distribution with
depleted values above the mean velocity. {\it Group 3} is the reverse with
positive skewness implying depleted values below the mean velocity.
The kurtosis values for both are a little high implying heavy tails.  The tail
index for {\it group 1} has a CN(0.20,3) distribution shape according to
Table 1 of \cite{BB93}, where 20\% of the points within a Gaussian width
of 3 $\sigma$ are bad.  In {\it group 3} the kurtosis value of 1.222 points
to a double exponential distribution. This would lead one to believe that the
groups are being pulled apart by the gravitational field of {\it group 2},
distorting their distributions.

	A histogram of velocities for {\it group 2} is presented in figure 3b.
All galaxies were within 3$\sigma$ of each other and the 
results for C$_{BI}$, S$_{BI}$, skewness, kurtosis,
asymmetry, and tail are shown in Table 3. The skewness and asymmetry
indices imply a Gaussian distribution. The kurtosis
index once again points to a tailed Gaussian distribution and the tail
index implies that we have a normal distribution. Further analysis on
{\it group 2} using 2D and 3D estimators demonstrate that these are probably
false positive rejections of a Gaussian distribution, see following subsections.

\subsubsection{cD velocity offset, Z$_{score}$}

	In relaxed clusters the cD galaxy should sit at the bottom of
the potential well, thus they should be at the center of the velocity
distribution (\cite{QL82}), which is borne by $\sim$70\% of cD clusters
(\cite{B94}).  A significant cD galaxy velocity offset from the cluster mean
(or the Z$_{score}$, \cite{GB91}) may be an indicator
of substructure.  \cite{B94} has shown that if one attempts to identify the
cD galaxy's host clump, using the methods in this paper to eliminate
non-cluster members, that most cases of ``speeding cDs'' disappear.

	In Table 4 the Z$_{score}$ values for {\it the cluster} and
{\it group 2} are presented.  One can see that the Z$_{score}$ for {\it group 2}
is slightly larger than that for {\it the cluster}, but
the Z$_{score}$ with bootstrapped errors bracket zero which
would {\it not} indicate a significant velocity offset, implying that
the Abell 119 cD is indeed sitting at the center of it's host clump.  The host
clump being {\it group 2}, which has a qualitatively similar Z$_{score}$
value. It can be seen that the fore and background clumps might tend to
balance out the Z$_{score}$ of {\it the cluster} because it is dependent
solely on the measured 1-D radial velocities of galaxies.

\subsubsection{2D and 3D tests}

	To further test {\it the cluster} and {\it group 2} data several
2 and 3 Dimensional substructure tests were used. The Lee statistic
(\cite{F88}) tests a 2D dataset for the presence of 2 equal sized groups
versus 1. The 3D \cite{DS88} $\Delta$ test, \cite{WB90} $\alpha$ test,
and \cite{B94} $\epsilon$ test all look for clumping in the
spatial {\it and} velocity data. The test results are
presented in Table 4. The Lee statistic results for {\it the cluster}
dataset imply a null result for the two group fit because of the low value of
L$_{RAT}$ and a p value of 117 (where p less than 25 implies a
statistically significant amount of substructure). On the other hand
a plot of the Lee statistic distribution (figure 5a) can help to define an
elongation axis, if any, of the 2D distribution. The highest point in
figure 5a defines the elongation axis of {\it the cluster} to be
$\phi_{max}$=86.4$\arcdeg$. Any multiple peaks seen in the Lee statistic
plot may be an indicator of more complex structure, even given
the low L$_{RAT}$, but there are no multiple peaks here.
The Lee statistic applied to {\it group 2} has a slightly
higher value of L$_{RAT}$,but the p value still rejects any
statistically significant substructure. The Lee statistic for
{\it group 2} plotted in figure 5b differs little from that of {\it the
cluster} in figure 5a. The elongation axis $\phi_{max}$=86.4$\arcdeg$ 
seen in {\it the cluster} plot of 5a (the peak in 5a) has vanished.

	In contrast all of the 3D tests (Table 4) report null for the
substructure hypothesis in {\it the cluster} data even though two in-falling
clumps along the line of sight (LOS) were found using KMM, but tests done by
\cite{B93} indicate that the 3D tests are insensitive to LOS mergers!

	2D and 3D tests were also applied to {\it group 2}. Since the
foreground and background groups were along the LOS of ``the cluster''
and are more or less evenly distributed in RA and DEC (see figure 4) one
would not expect to see much of a difference in comparison with {\it the
cluster} data.  The Lee statistic for {\it group 2} has a slightly higher
value of L$_{RAT}$, but the p value still rejects any statistically
significant substructure. The Lee statistic result for {\it group 2} is plotted
in figure 5b. It does not differs greatly from that of {\it the cluster}
dataset showing the elongation angle $\phi_{max}$=82.8$\arcdeg$.

	Of the three 3D tests only the $\Delta$ test statistic gave a positive
rejection of the Gaussian hypothesis. This may be a false positive since no
other estimators gave the same result and, as pointed out by \cite{B93}, the
$\Delta$ test is the more optimistic of the three.
This result coupled with a low L$_{RAT}$ for {\it group 2}
from the Lee statistic lends support to the 3D tests null result.

\subsubsection{Rotation}

	As pointed out by \cite{M92} a smooth gradient
in the velocity field may complicate use of the $\Delta$ statistic by giving
a false positive substructure result. Even though in Abell 119 the $\Delta$
statistic has a null hypothesis for {\it the cluster} data set one is still
interested in knowing whether clusters in general show signs of
rotation.  An estimate of cluster rotation can be made by calculating a
binned C$_{BI}$ along the elongation axis (as defined by the Lee statistic).
Figure 6a shows C$_{BI}$(R)-C$_{BI}$(global) versus R with 90\% bootstrapped
confidence intervals. There is no clear gradient in the data and therefore
implies little in the way of cluster rotation. There is however a strong
discontinuity at radii of around 1 h$^{-1}$Mpc. This {\it could} just be a
sampling effect. If one looks at the original positions as measured
from the ESO plates (figure 1a) one cannot help but see two ``voids'' in the
lower left hand corner and upper right hand corner at about 1 h$^{-1}$Mpc
in radius.  This also manifests itself in the measured velocities
(figure 4).

	For {\it group 2} (figure 6b) much the same situation as above
is found. No clear gradients seem to exist, although the discontinuous feature
at $\sim$1 h$^{-1}$Mpc persists, as expected. This result
would tend to support the $\Delta$ test finding of substructure in
{\it group 2} given a lack of evidence for any strong velocity field gradients.

\subsubsection{Velocity Dispersion Profile (VDP)}

	Variations in the velocity dispersion with radius may indicate
a condition of non-equilibrium (\cite{K87}). To test this for ``the cluster''
figure 7a plots radius versus velocity.  The caustics for ``the cluster''
data are well defined except for three points at the top of figure 7a.
Eliminating these three points (which are actually part of the background
group picked with KMM) a plot of the cumulative S$_{BI}$ (velocity dispersion)
versus radius is shown in figure 8a.
As one can see in figure 8a, the velocity dispersion
falls with radius as is seen with many well studied rich clusters 
(e.g. Abell 3266 in \cite{QRW96},  and others in \cite{HK96}).

	For {\it group 2} radius versus velocity is show in figure 7b and
cumulative velocity dispersion versus radius in figure 8b. Now the situation 
has changed dramatically. Both plots show a roughly flat distribution,
even out to large radii. Either the fore and background groups have been 
eliminated incorrectly (which is not supported by the previous work of
\cite{FAB93} nor the X-ray data) or one is witnessing the effects of velocity
anisotropies in the central region (\cite{F96}) where the effects of
dynamical friction may be slowing down the more luminous central galaxies.
One may recall that the frictional force is proportional to the local matter
density (\cite{C43}) which is higher in the central region of a cluster.
As well, \cite{HK96} agree that it may be anisotropic projection effects
that cause inverted VDPs. {\it Group 2} does not have an inverted VDP, but
it may help to explain the flat VDP seen.

\subsubsection{Rosat versus AKM}

	Using the adaptive kernel map (AKM) first applied by \cite{Beers91}
to the 2-D galaxy distribution one can attempt to make a comparison between the
contours generated from the number density of galaxies deemed to be
in {\it group 2}, and that from the X-ray density contours of Rosat.

	Figures 9a and 9b show the Rosat and AKM contour optical
overlays for a 1.5x1.5 h$^{-1}$Mpc region centered on the central cD. The
data is restricted to the inner 1.5 h$^{-1}$Mpc since at this redshift that
is the extent to which one can gain meaningful information from the X-ray data.

	The Rosat X-ray data in figure 9b has been smoothed with a 2 pixel FWHM 
Gaussian using the imsmooth task in the PROS\altaffilmark{1} X-ray reduction
package.  The image was obtained from the publically released
HEASARC Rosat CD Volume 2 (\cite{C94}).

\altaffiltext{1}{PROS is developed, distributed, and maintained by the
Smithsonian Astrophysical Observatory, under partial support from NASA
contract NAS5-30934}

	Since both the AKM and Rosat data presented in figures 9a and 9b
cover the same 1.5x1.5 h$^{-1}$Mpc region a direct, albeit qualitative,
comparison of the matter density (Rosat/X-ray) to the 2-D galaxy number density
can be made. It is obvious that a NNE elongation in the central
regions of both plots manifests itself and in fact coincides with the
elongation axis {\it objectively} obtained using the Lee test statistic.
This strong evidence suggests that if one obtains enough galaxy redshifts
in a cluster one can accurately begin to estimate the local matter
density with confidence.

\subsubsection{Radio}

	The core of Abell 119 has been radio mapped at 20cm with the VLA by
\cite{ZBO89}. They claim the elongation of the cD may indicate a Wide
Angle Tailed source. The NNW elongation seen in the radio mapped cD does
not correspond to the large scale NNE elongation seen in this cluster.
If the cluster had formed recently one might
expect to find the radio structure mimicking the larger scale structure,
but since this is not the case it is presented as evidence that {\it group 2}
was not recently formed, and since one would expect more substructure with
younger systems it further supports the lack of substructure seen
in the velocity field.

\subsubsection{Mass}

In Table 5 the mass estimates of the {\it three groups} are
reported. The virial, average, mean and projected mass estimators as described
in \cite{HTB85} were used.  The mass estimate within 0.5 h$^{-1}$Mpc for
{\it group 2} is reported so as to compare with the Rosat X-ray estimate
of \cite{J96}. Good agreement within the errors is found for the average and
median mass estimators.  The mass for R$<$2.3Mpc is also shown
(the limit of this survey) as a rough comparison with the X-ray estimate of
\cite{AK83} (Table 5) who reach a radius of 1.93 h$^{-1}$Mpc and whose value is far above
the higher error bar on all four mass estimators determined from the data in
this paper. This is likely due to the fact that \cite{AK83} used a $\beta$
(the value of the dimensionless temperature)
of 1, whereas other studies (\cite{JF84}) have since pointed to values between
0.5 and 0.7 for most clusters of galaxies.

In Table 5 calculated masses are also shown for subgroups 1 and 3. As noted
in section 5.1.1 these groups are not likely to be virialized.
This is because they are being tidally disrupted by the gravitational
field of the main cluster group which would distort their distribution and
prevent one from obtaining an accurate estimate of the mass using
the virial theorem. Nonetheless the numbers are printed here for
comparison with any future estimates.


\subsection{Abell 133}

\subsubsection{1D tests}

	ROSTAT and 3$\sigma$ iterative clipping were
employed to keep 120 velocities in the range 15279$< v <$18846 km sec$^{-1}$.
Table 3 shows the 90\% confidence intervals about the location (C$_{BI}$)
and scale (S$_{BI}$) of the 3 S$_{BI}$ clipped data using 10000 bootstraps.

	Table 3 also presents the results of the shape estimators on
the velocity distribution (see figure 10a). Of the 4 estimators
only the large kurtosis value would lead one to believe the distribution
is non-Gaussian. The kurtosis
implies the distribution is heavily tailed and that one should run
KMM to look for in-falling groups along the 
line of sight. Attempts to identify two, three, four and five groups
all failed with large margins.  No multi-group fit came back with a 
null rejection of the single Gaussian hypothesis.

Given the failure of KMM to discern any multiple Gaussian structure one must
look to the Z$_{score}$, 2D and 3D tests for any confirmation of the kurtosis.

\subsubsection{cD velocity offset, Z$_{score}$}

	Table 4  shows the Z$_{score}$ and cD peculiar velocity
for Abell 133. There is a case for a ``speeding cD'' given the fact that the
Z$_{score}$ value with error does {\it not} bracket zero. Given that no
possible host subclumps have been objectively verified one must assume this
an indicator of dynamical youth. See the VDP section below for more.

\subsubsection{2D and 3D tests}

	Table 4 also presents the 2D and 3D results using the 120 galaxies
within 3S$_{BI}$. Note that the centroid of the galaxy positions was taken
as the center of the cluster. This was justified
by the Z$_{score}$ value indicating the cD is not at the center
of the cluster, and therefore not a good place to pick the cluster center.
It is important to pick a good center as some of the substructure indicators
are sensitive to this value.

	The Lee statistic has a small p value of 15. This indicates that a
two group fit versus one is likely. Recall again that a p value of less
than 25 indicates a statistically significant probability.
Figure 10b shows a plot of the Lee distribution with a peak at 93.6$\arcdeg$.

The lack of multiple peaks
and the high value of L$_{RAT}$ (Table 4) continue to insist that no more
than two groups are likely in the X-Y plane. The Lee statistic was
also run on the inner 1.5 $h^{-1}$Mpc region so as to compare with the
AKM and Rosat data below. There were two peaks in the Lee distribution
implying more complex structure as mentioned. One peak was at 88$\arcdeg$
which corresponds to the elongation seen in the AKM and Rosat maps (see
section 5.2.6).

	The other 3D estimators should verify this 2-D structure if it
exists. While these estimators are not proficient at LOS substructure (which,
outside the Kurtosis and Lee results, the 1D tests and KMM failure have
ruled out) they
are sensitive to 2D structure in the plane of the sky. Table 4 contains
the values for the $\Delta$, $\alpha$, and $\epsilon$ tests. Only the $\Delta$
statistic indicates a statistically significant level of substructure, in
keeping with the Lee statistic. However, again it should be kept in mind that 
the $\Delta$ is the most optimistic, or most likely to give a false
positive of the three.

\subsubsection{Rotation}

	Figure 11a shows the C$_{BI}$ velocities and their 90\% confidence
intervals binned along the elongation axis
as given by the Lee statistic. There are no clear gradients which would
indicate rotation. 

This further emphasizes the $\Delta$ statistic result above. If one did see
signs of rotation this might show up as a positive substructure result
in the $\Delta$ test where small values of V$_{rot}$/$\sigma$ may cause
detection by the 3-D diagnostics. Since there are no signs of rotation
in C$_{BI}$ one can put more confidence in the $\Delta$ statistic result.

However there are discontinuities at -1 and 1.25 h$^{-1}$Mpc.
The discontinuity at -1 may be explained by the low number of galaxies in the
last two bins and the lack of sample south of 1.5 h$^{-1}$Mpc.

\subsubsection{Velocity Dispersion Profile (VDP)}

	As for Abell 119 variations in the velocity dispersion were tested by
plotting velocity versus radius (the caustics) in figure 11b and cumulative
S$_{BI}$ in figure 11c.  Both plots are fairly flat out to 2.5 h$^{-1}$Mpc
as in {\it group 2} of Abell 119. The flat VDP within 1 h$^{-1}$Mpc is again
an indicator of galaxy velocity anisotropies and is supported by
our Z$_{score}$ result for Abell 133.

\subsubsection{Rosat versus AKM}

	Figures 12a and 12b show the AKM and Rosat contour optical
overlays for a 1.5x1.5 h$^{-1}$Mpc region centered on the central cD. These are
once again restricted to the inner 1.5 Mpc since that is the extent to which
one can gain meaningful information from the X-ray data at this distance.

	The Rosat X-ray data has been smoothed with a 2 pixel FWHM 
Gaussian using the imsmooth task in the PROS X-ray reduction package.
As above, the image was obtained from the publically released
HEASARC Rosat CD Volume 2 (\cite{C94}).

	If one focuses on the inner regions one can discern a slight NNE SWW
elongation of the X-ray gas and galaxy distribution. As one goes farther
away from the center the contours push out to the SE in both maps. 
This is in agreement with the elongation seen in the Lee statistic
result for data within 1.5 h$^{-1}$Mpc. As well there appears to be
small structures, again supported by the flat VDP.

\subsubsection{Radio}

Abell 133 has been reported as a strong radio source and studied by several
groups (\cite{S89}, \cite{OWG93}, \cite{G94}). The radio structure of the
cD has been resolved into two sources by \cite{OWG93}(figure 1), but the
orientation of the double structure does not correspond to the elongation axis
of the cluster. The wide area radio map by \cite{S89}(figure 9) also resolves
the multiple cD components, but otherwise has no correspondence with the
elongation angle of the cluster. As for Abell 119 if the cluster had formed
recently one might
expect to find the radio structure mimicking the larger scale structure,
but since this is not the case it is presented as evidence of an older
system with little substructure.

\subsubsection{Mass}

	Table 5 compares this paper's mass estimate for Abell 133
with the X-ray mass estimates
of \cite{J96} for R$<$0.5h$^{-1}$Mpc and R$<$1.5h$^{-1}$Mpc.
For R$<0.5h^{-1}$Mpc all four mass estimates are in agreement with the
X-ray data within their respective 90\% confidence intervals.
For R$<$1.5h$^{-1}$Mpc of the four estimators only the projected mass estimator
(PME) does not overlap with the X-ray value. This is quite a surprising result
given the substructure seen in the Rosat image in combination with the
Lee, $\Delta$, Z$_{score}$, and VDP results suggesting a system non-ideal
for virial estimates of mass.

\section{Discussion}

	Starting with 174 galaxies in Abell 119, which include this paper's
newly recorded
velocities and those obtained from the literature, 3 sigma iterative clipping
left 155. From the 1-D velocity distribution the ROSTAT statistical program
yielded a high positive kurtosis in the remaining 155 galaxies pointing to
tails heavier than expected for a Gaussian distribution. Subsequently
the KMM partitioning algorithm
was used to search for overlapping Gaussian distributions in the velocity
field. Three overlapping distributions were found rejecting the single Gaussian
distribution at the 97.9\% confidence level. A main group of 125 members
and two smaller groups of 11 and 17 were found. Further 1-D analysis with ROSTAT
on the main subgroup of 125 members ({\it group 2}) lent support to a Gaussian
distribution of velocities, while this was not the case for the 2 smaller sub
groups.  As well, the central galaxy was not significantly offset from the
mean velocity of the cluster in {\it group 2} implying the lack of a speeding
cD.  The 2-D Lee statistic did not detect two groups in {\it group 2}, but
did lend support to an elongation axis near 82.8$\arcdeg$. Of the 3-D tests used
on {\it group 2} only the Delta test was positive, but it should be noted that
it is the more likely to detect a false positive of the 3-D estimators.
No clear gradients were found in the velocity field of {\it group 2}, further
supporting the delta statistic which is susceptible to such gradients.
The flat velocity dispersion profile for {\it group 2} points to velocity
anisotropies in the central region.  Rosat X-ray data was compared with the
number density of galaxies in {\it group 2} as plotted using an Adaptive Kernel
Map (AKM) within 1.5h$^{-1}$Mpc of the central cD galaxy. The elongation pointed
out with the Lee statistic is replicated in both the Rosat and AKM maps implying
that if one obtains enough galaxy redshifts in a cluster one can accurately
begin to estimate the local matter density with confidence.
Virial mass estimates of \cite{J96} from Rosat data compare nicely
with those obtained from the velocity data presented here for those
galaxies within 1.5h$^{-1}$Mpc of the central galaxy of {\it group 2}.

	In Abell 133 a dynamical analysis using newly collected velocity data
in combination with that of the literature was done starting with 153
velocities.  3 sigma iterative clipping reduced the 153 to 120.
The remaining 1-D velocity data was analyzed with ROSTAT to yield a relatively
high positive kurtosis, but when KMM was used to search for multiple
Gaussian fits none were found with a significance level higher than
that of a single Gaussian. However, the central cD galaxy was significantly
offset from the cluster mean velocity which is an indicator of
dynamical youth especially since there appears no host subclump for the
central cD galaxy. This is further supported by the flat
velocity dispersion profile seen in the inner part of the cluster implying
velocity anisotropies.  The Lee statistic was positive for a 2 group fit,
but of the 3-D statistical indicators only the delta statistic supported the
finding. However the delta statistic result itself was
further supported by the lack of velocity gradients in the C$_{BI}$ 
velocities along the elongation axis found with the Lee statistic.
The Lee statistic run on the inner 1.5h$^{-1}$Mpc gives an elongation axis
at 88$\arcdeg$, which corresponds to what is seen in the Rosat and AKM images.
The Lee statistic run on this portion of the data does show multiple
peaks indicating more complex structure, also supported by what
is seen in the Rosat image.
The complex structure found above is again bolstered by the
flat VDP seen, indicating velocity anisotropies in the inner regions
of the cluster.  All 4 mass estimates agree with the exception of the one for
the projected mass estimator within R$<$1.5h$^{-1}$Mpc. This is quite a
surprising result given the substructure seen in the Rosat image in
combination with the Lee, $\Delta$, Z$_{score}$, and VDP results which
suggest a system non-ideal for virial estimates of mass. This can
only be disentangled in the future by further velocity measurements in
the field of this cluster. 

	Both of these X-ray clusters seem to demonstrate virialization to
large radii given the good correlation between the X-ray and velocity mass
estimators. However the Lee statistic for both of these clusters points to an
elongation of the cluster. In Abell 133 a hint at two groups in the plane of
the sky is also observed.  In Abell 119 the elongation is less 
pronounced than that of Abell 133 where within 1 h$^{-1}$ Mpc the distribution
of galaxies gives one the qualitative impression of a small group falling
toward the cD galaxy. The X-ray gas in this region of Abell 133 is also
elongated in the center as can be seen in figure 12b. The situation for
Abell 133 seems much more clear than that of Abell 119. Most of the
indicators point to substructure in Abell 133, whereas in Abell 119
the results are more mixed. In either case it is clear that when
substructure is taken into account via the statistical methods
demonstrated within this paper the velocity+spatial versus
X-ray virial estimators can compare nicely.


\acknowledgements 
	The authors would like to thank the Director of the Observatories of
the Carnegie Institution of Washington for generous allocation of telescope
time at Las Campanas. It is a pleasure to thank Steve Shectman for use of
the spectrograph and P. Harding and
John Filhaber for help with instrument setup. We would also like to thank
Ricardo Flores and Tina Bird for useful discussions and the use of various
programs. This research has made use of the NASA/IPAC Extragalactic Database
(NED) which is operated by the Jet Propulsion Laboratory, California Institute
of Technology, under contract with the National Aeronautics and Space
Administration. As well, This research has made use of data obtained
through the High Energy Astrophysics Science Archive Research Center Online
Service, provided by the NASA/Goddard Space Flight Center.  This project was
partially supported by FONDECYT grants 1960413 and 8970009.  M.J. Way has been
supported by Research and Research Board awards at The University of
Missouri--St.Louis.



\newpage

\figcaption[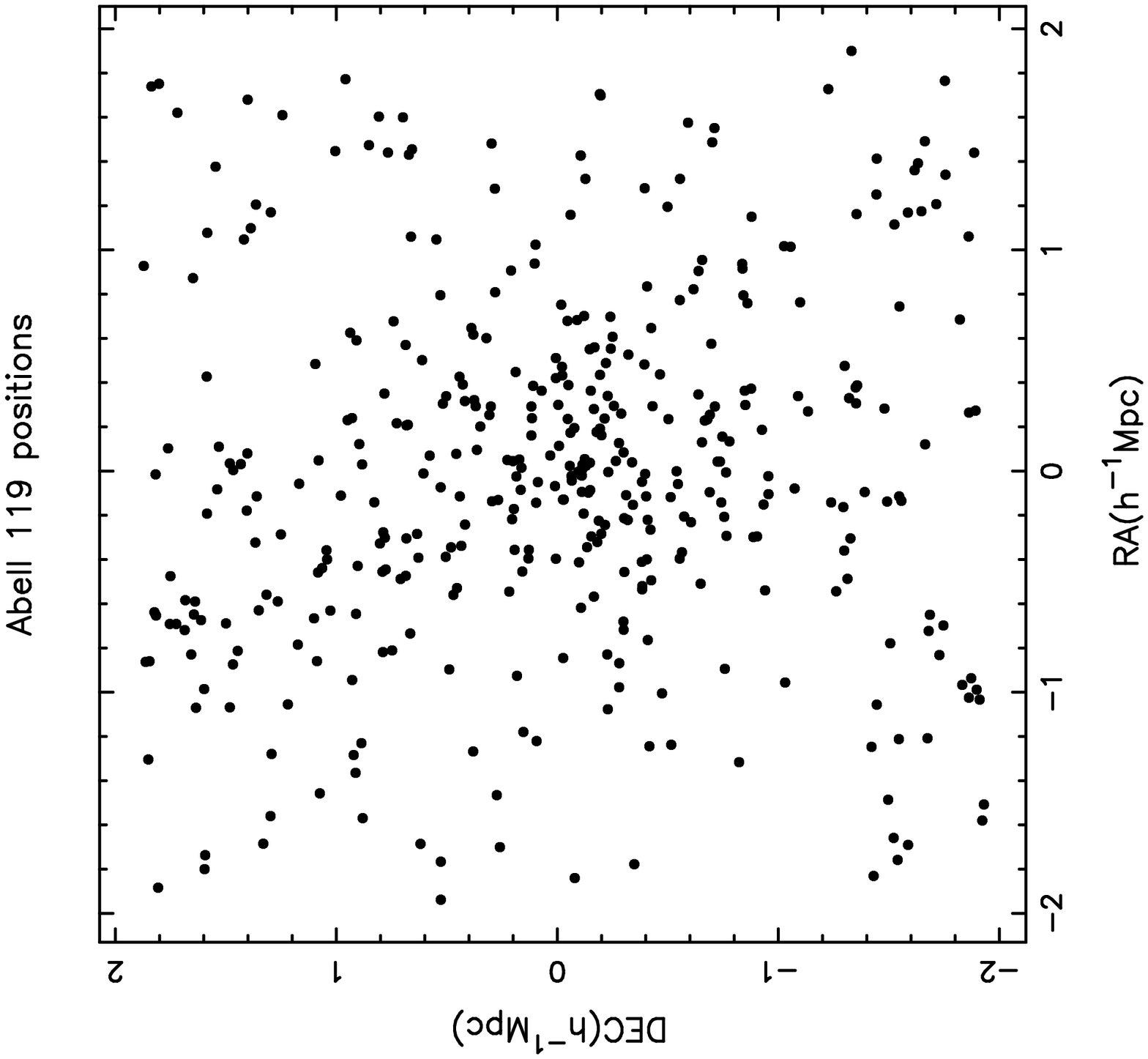]{X-Y positions for galaxies in Abell 119 picked from
the ESO Quick Blue survey plates.
\label{fig1a}}

\figcaption[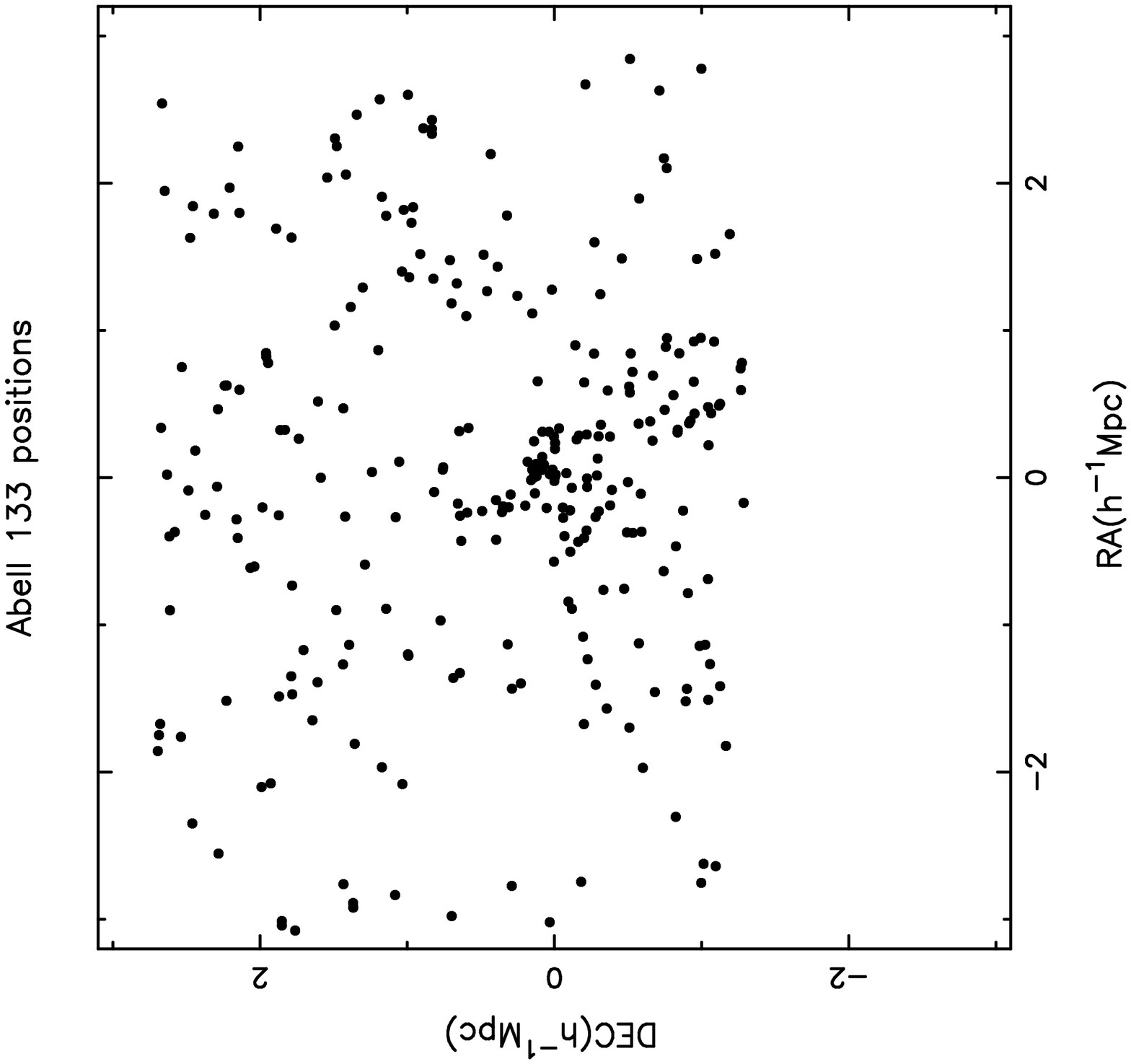]{X-Y positions for galaxies in Abell 133 picked from
the ESO Blue plates.
\label{fig1b}}

\figcaption[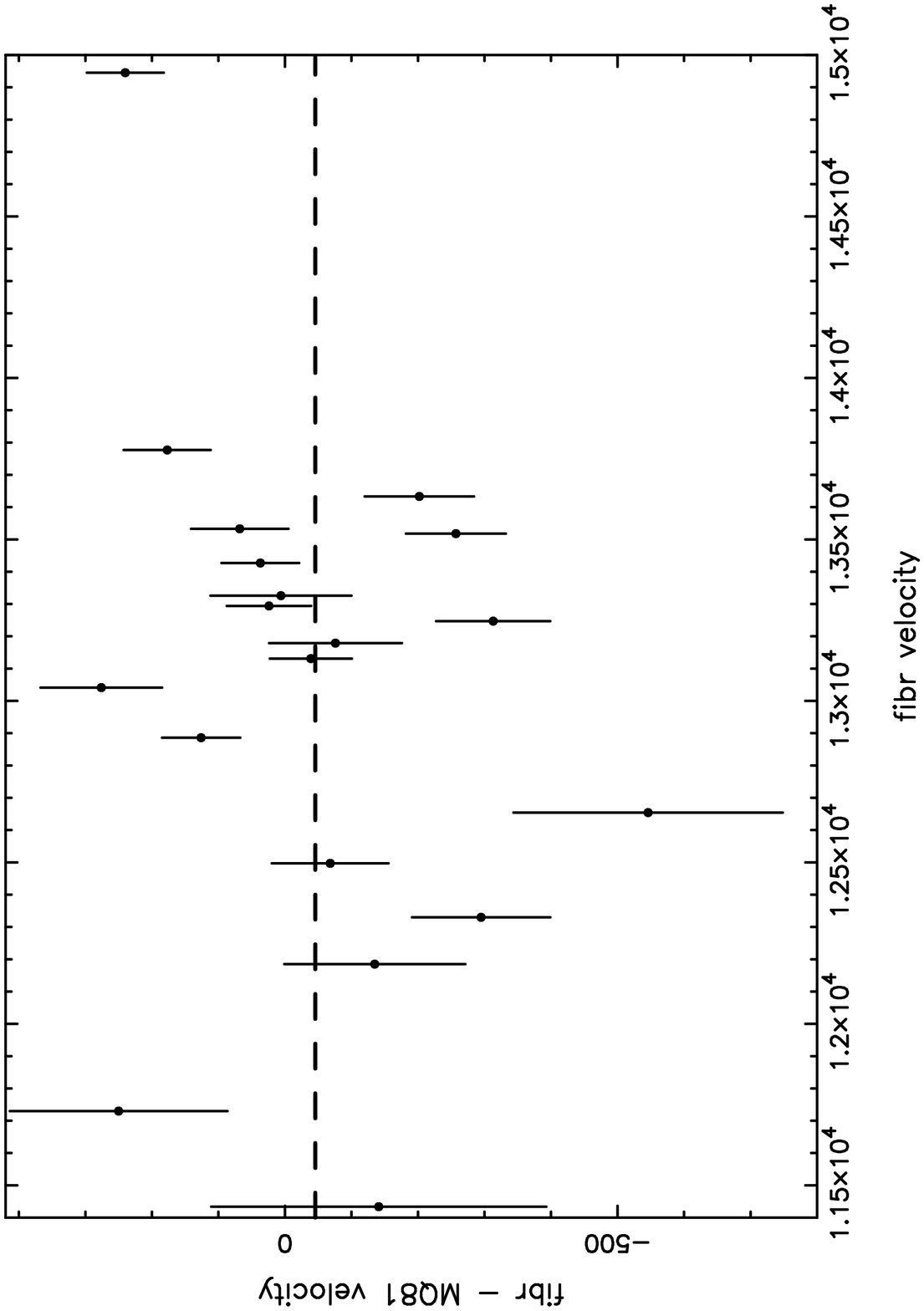]{Velocity residuals between Abell 133 and 119 in
this paper and \cite{MQ81}. A 0th order Polynomial line was fit for those values
within 3.0 $\sigma$ of the mean.
\label{fig2a}}

\figcaption[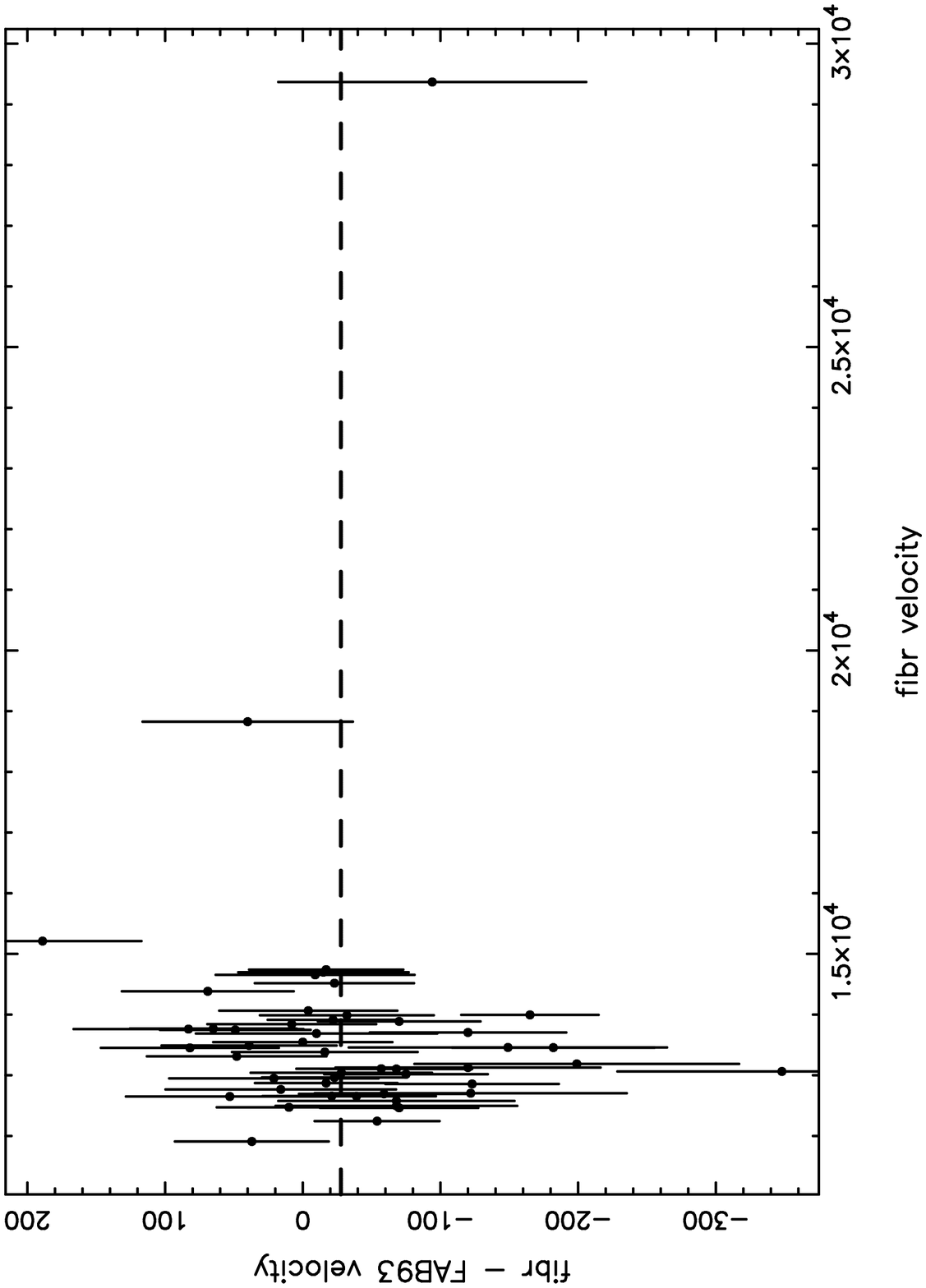]{Velocity between Abell 133 and 119 in this paper and
\cite{FAB93}. Same line fitting technique as used in Figure 2a.
\label{fig2b}}

\figcaption[way3a.ps]{Velocity histogram of Abell 119 for the 153 galaxies
within 3S$_{BI}$.
\label{fig3a}}

\figcaption[way3b.ps]{Velocity histogram of Abell 119 Group 2.
\label{fig3b}}

\figcaption[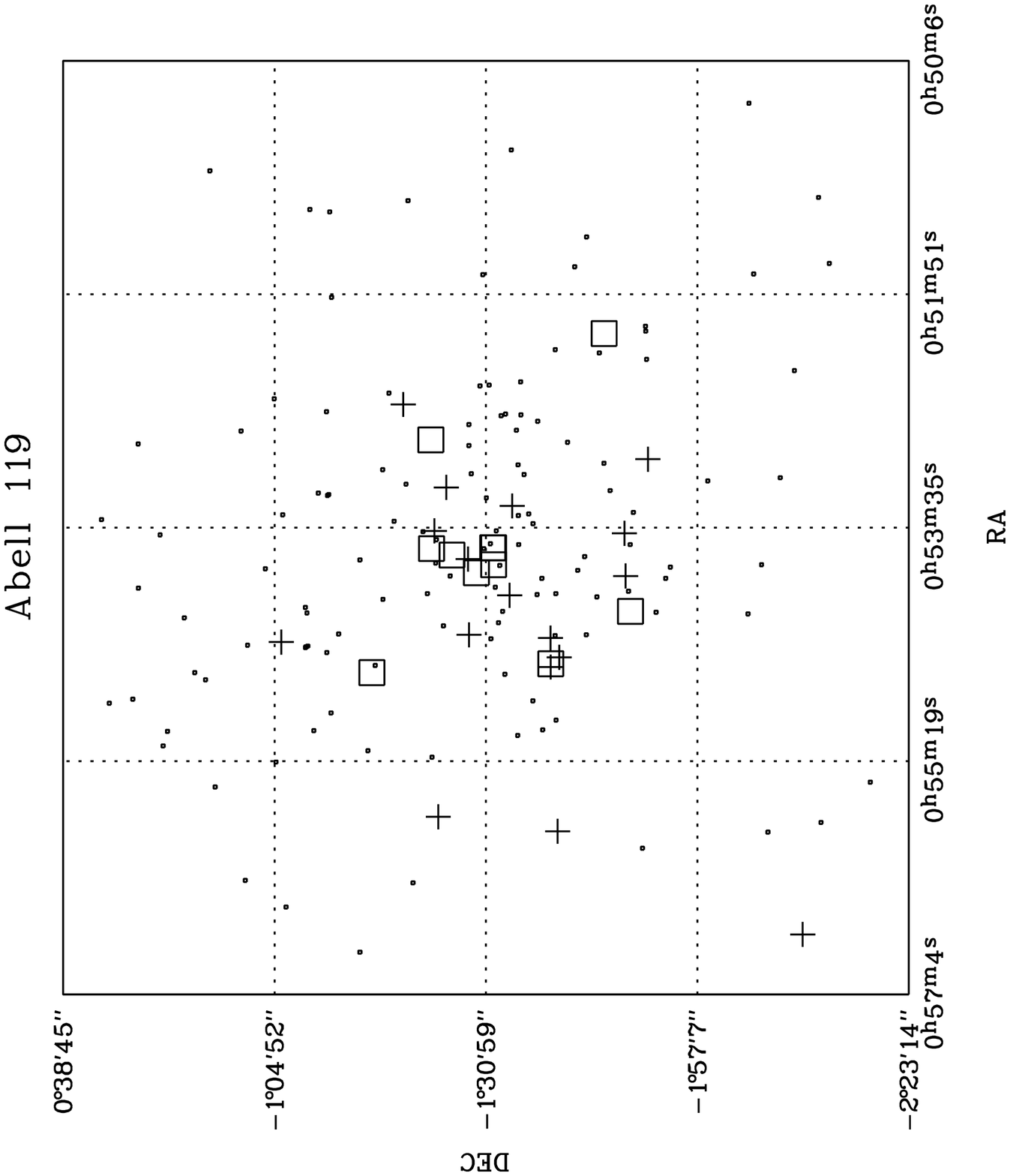]{Three groups from the 3S$_{BI}$ clipped
Abell 119 data partitioned with KMM.  Group 1 is denoted by $\Box$ (11 members),
group 2 by $\bullet$ (125 members) and group 3 by + (17 members).
\label{fig4}}

\figcaption[way5a.ps]{Lee statistic for Abell 119, LRAT = 2.231374,
N$_{gal}$= 153.
\label{fig5a}}

\figcaption[way5b.ps]{Lee statistic for Abell 119 group 2, LRAT = 1.5829,
N$_{gal}$= 125.
\label{fig5b}}

\figcaption[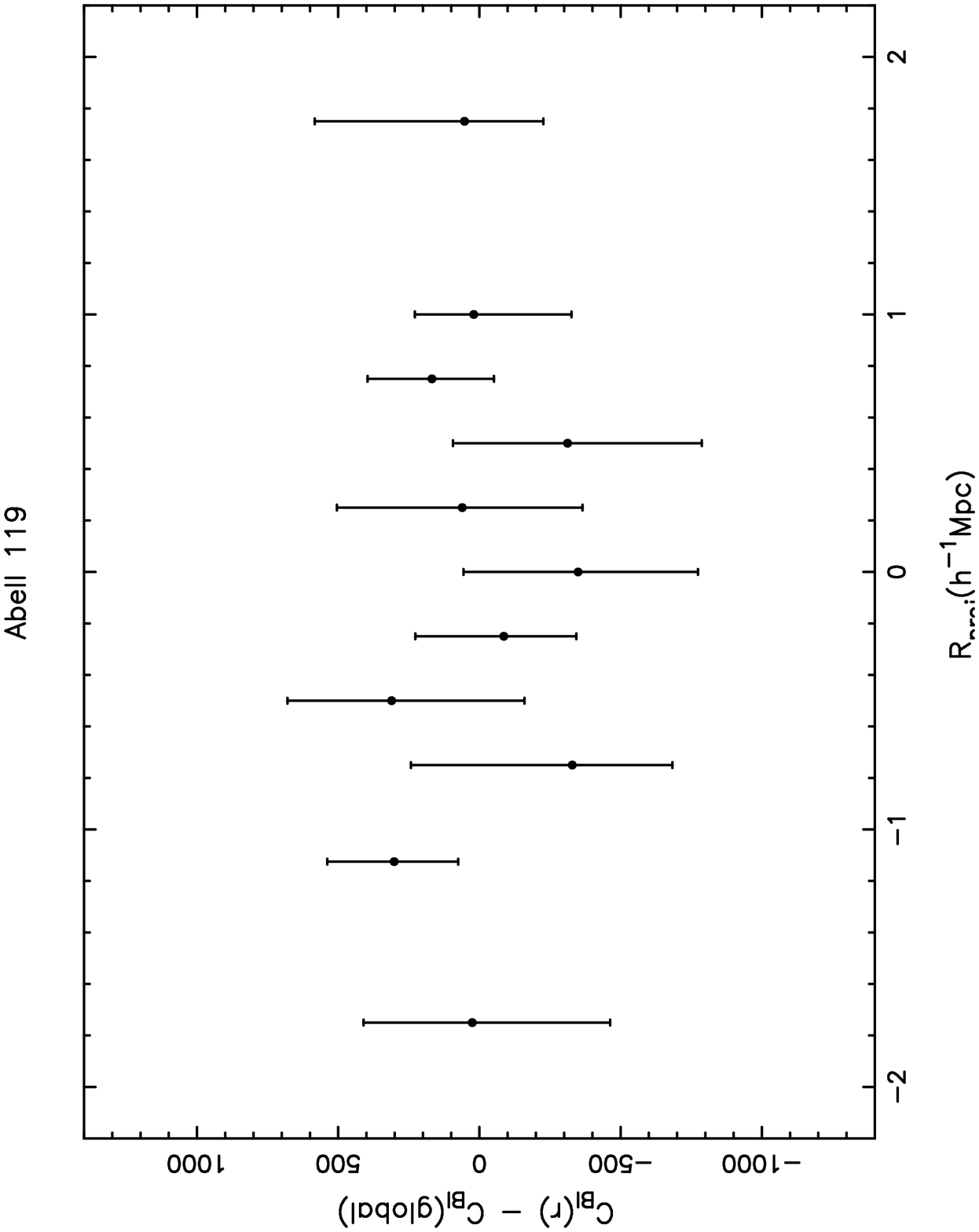]{Abell 119 C$_{BI}$(R)-C$_{BI}$(global) versus R. There is
no clear gradient in the data implying a lack of any clearly definable rotation.
\label{fig6a}}

\figcaption[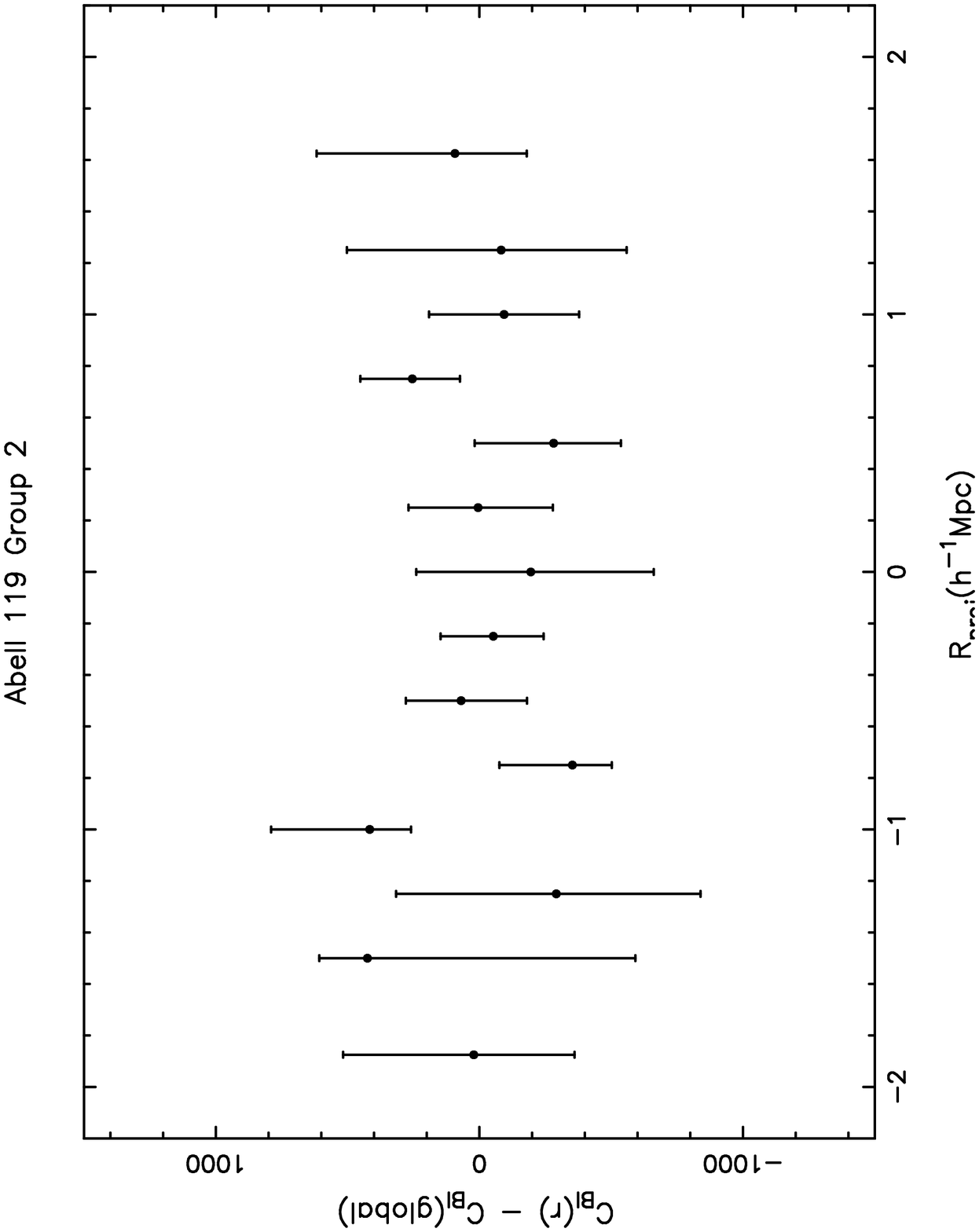]{Abell 119 Group 2 C$_{BI}$(R)-C$_{BI}$(global) versus R.
Once again there is no clear gradient in the data implying a lack of any
clearly definable rotation.
\label{fig6b}}

\figcaption[way7a.ps]{Caustics for Abell 119 (153 galaxies).
\label{fig7a}}

\figcaption[way7b.ps]{Caustics for Abell 119 Group 2 (125 galaxies).
\label{fig7b}}

\figcaption[way8a.ps]{Abell 119: Binned cumulative S$_{BI}$ versus R.
(153 galaxies).
\label{fig8a}}

\figcaption[way8b.ps]{Abell 119 Group 2: Binned cumulative S$_{BI}$ versus R.
(125 galaxies).
\label{fig8b}}

\figcaption[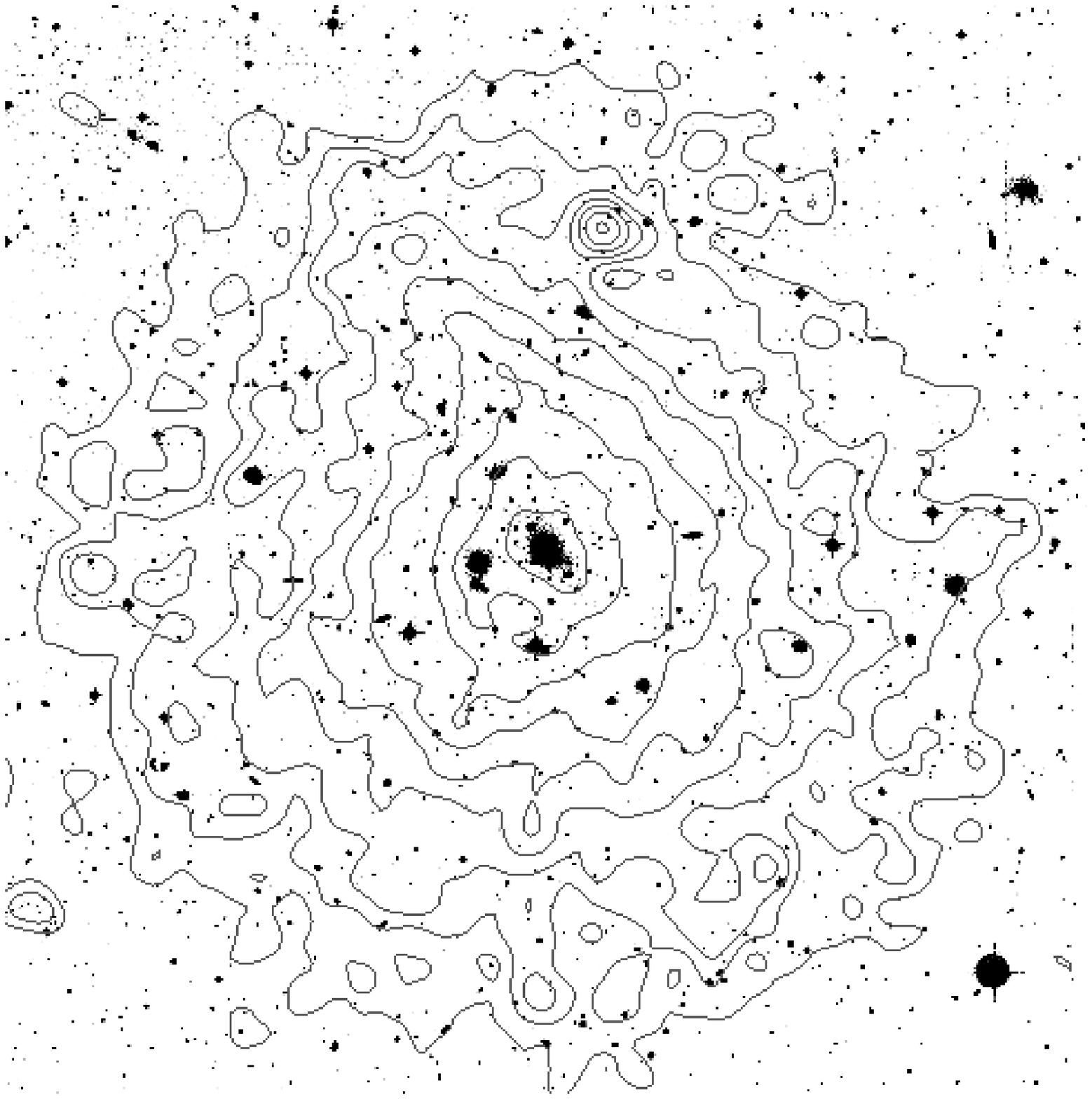]{Abell 119 Rosat X-ray - optical map overlay within 
R$<$1.5h$^{-1}$Mpc.
\label{fig9a}}

\figcaption[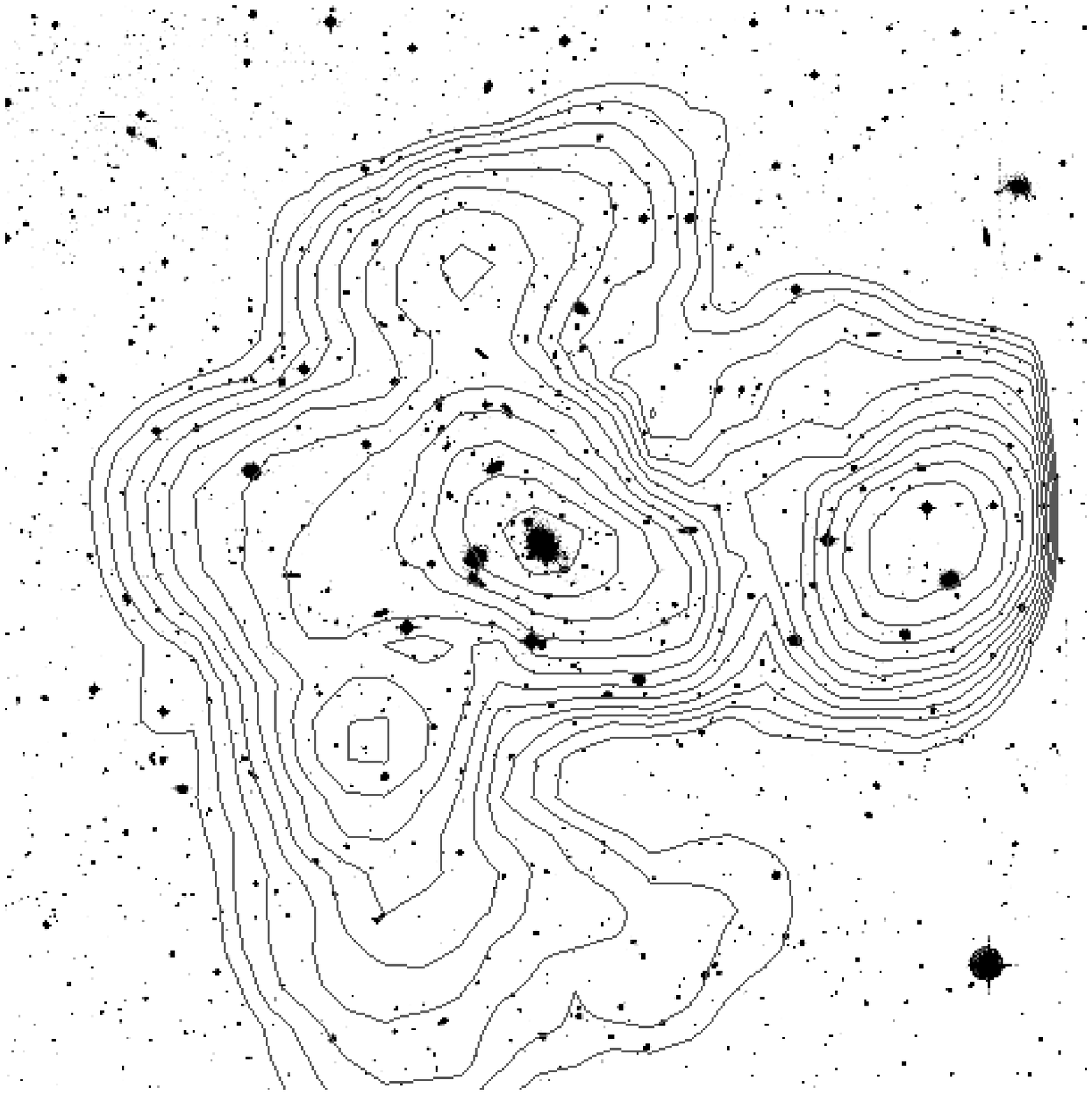]{Abell 119 Group 2 Adaptive Kernel - optical map overlay
within R$<$1.5h$^{-1}$Mpc.
\label{fig9b}}

\figcaption[way10a.ps]{Abell 133 Velocity Histogram for the 120 galaxies
within 3S$_{BI}$.
\label{fig10a}}

\figcaption[way10b.ps]{Lee statistic for Abell 133. LRAT = 2.415
\label{fig10b}}

\figcaption[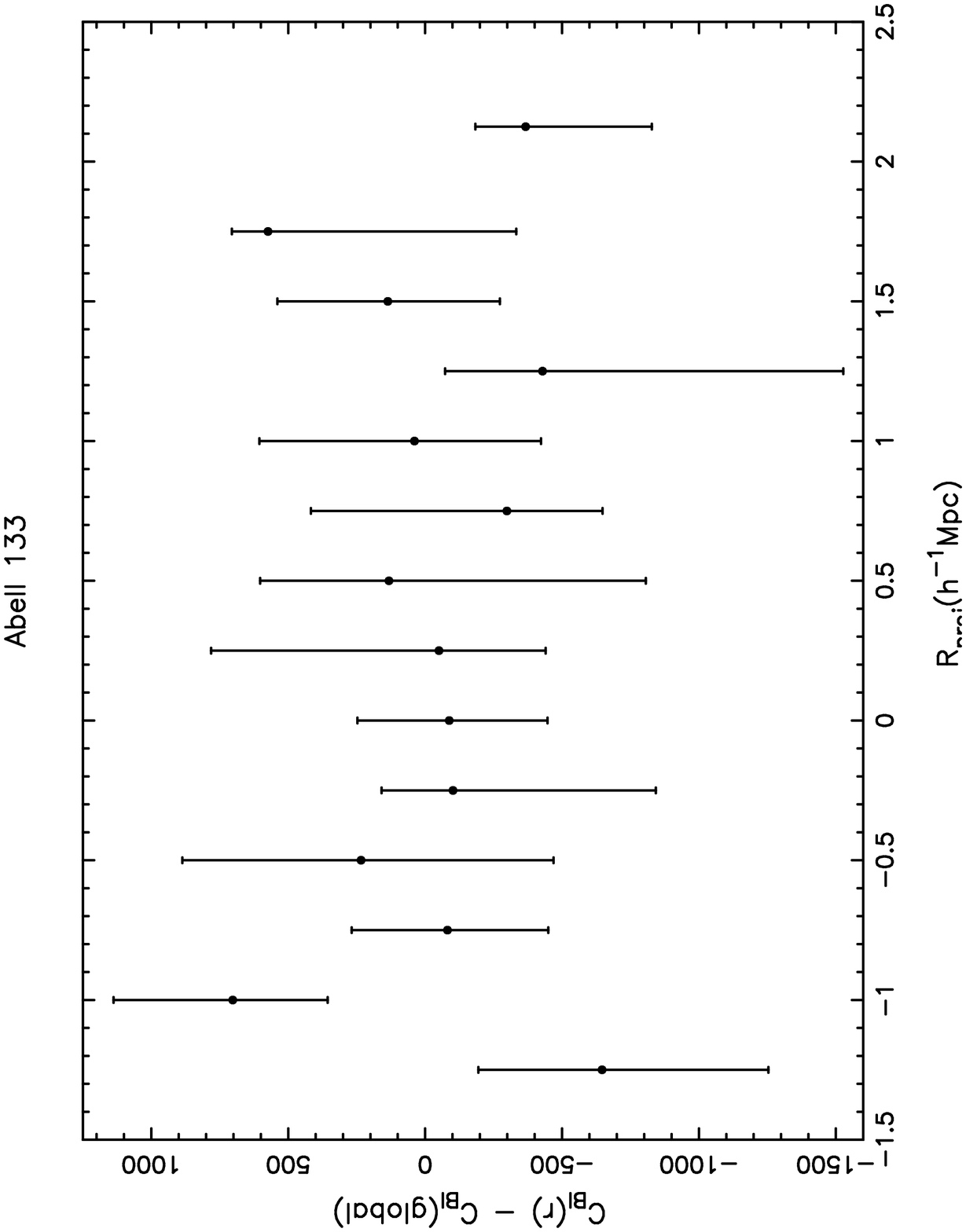]{Abell 133 C$_{BI}$(R)-C$_{BI}$(global) versus R. There is
no clear gradient in the data implying a lack of any clearly definable rotation.
\label{fig11a}}

\figcaption[way11b.ps]{Caustics for Abell 133.
\label{fig11b}}

\figcaption[way11c.ps]{Abell 133: Binned cumulative S$_{BI}$ versus R.
\label{fig11c}}

\figcaption[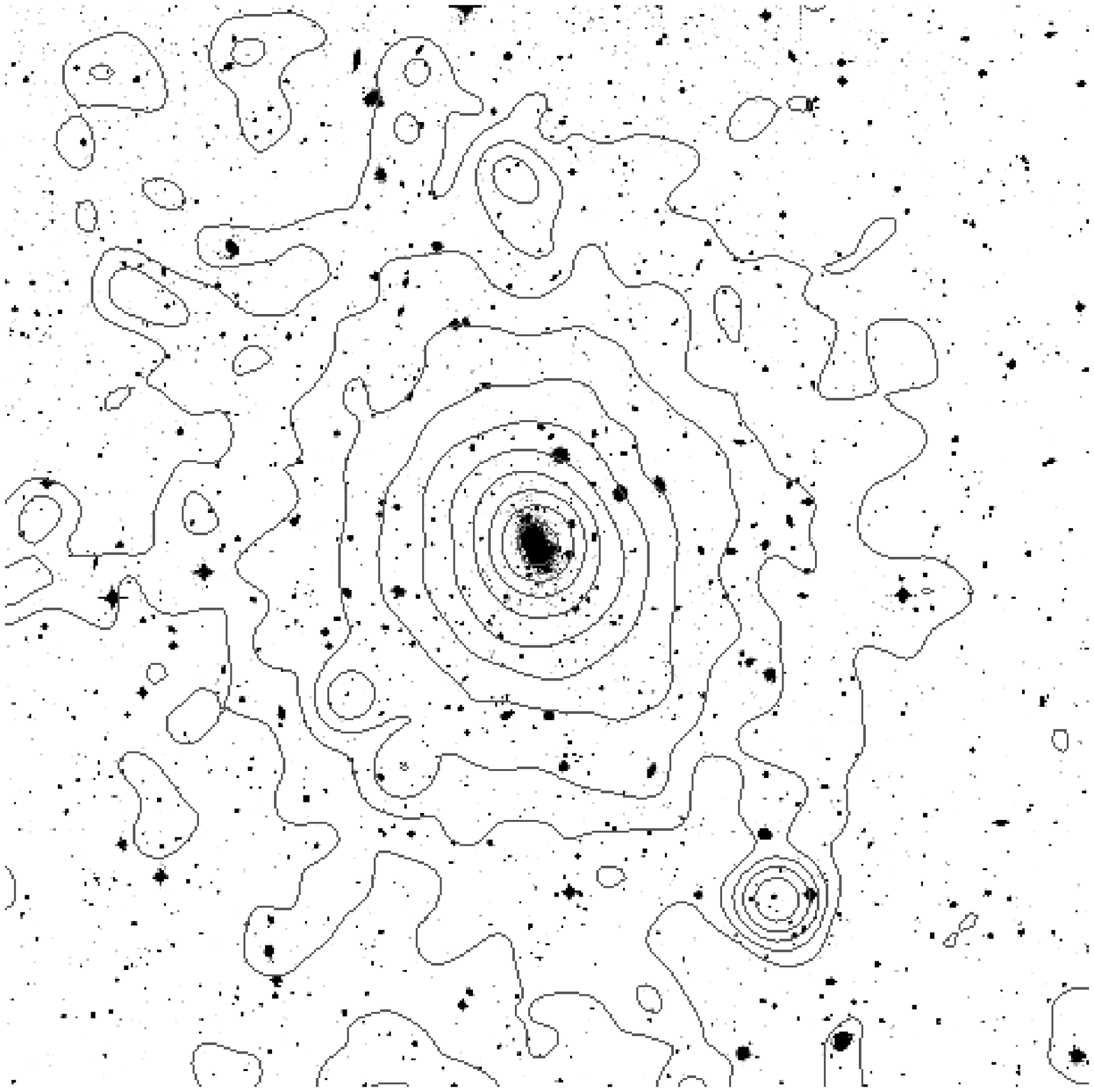]{Abell 133 Rosat X-ray - optical map overlay within 
R$<$1.5h$^{-1}$Mpc.
\label{fig12a}}

\figcaption[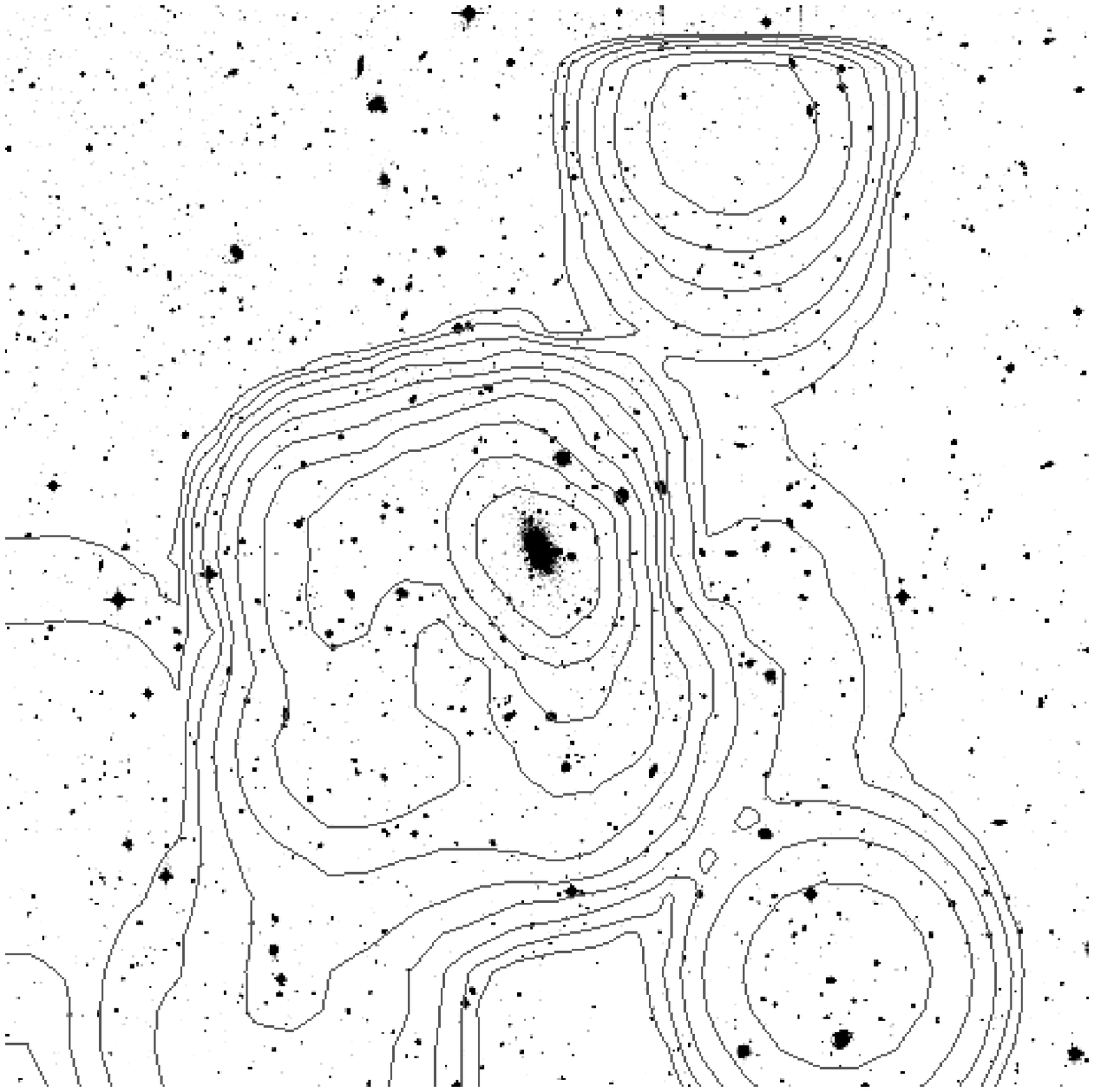]{Abell 133 Adaptive Kernel - optical map overlay
within R$<$1.5h$^{-1}$Mpc.
\label{fig12b}}


\begin{thebibliography}{}

\bibitem[Abramopoulos \& Ku 1983]{AK83}
Abramopoulos, F. \& Ku, W. 1983, \apj ~271, 446

\bibitem[Ashman et al. 1994]{A94}
Ashman, K.M., Bird, C.M., \& Zepf, S.E. 1994, \aj ~108, 2348

\bibitem[Beers et al. 1990]{BFG90} 
Beers, T.C., Flynn, K., \& Gebhardt, K. 1990, \aj ~100, 32

\bibitem[Beers et al. 1991]{Beers91} 
Beers, T.C., Forman, W., Huchra, J.P., Jones, C., \& Gebhardt, K. 1991,
\aj 102, 1581

\bibitem[Bird 1994]{B94}
Bird, C. 1994 \aj ~107, 1637

\bibitem[Bird 1993]{B93}
Bird, C. 1993, Ph.D thesis, University of Minnisota and
Michigan State University

\bibitem[Bird \& Beers 1993]{BB93}
Bird, C., \& Beers, T.C. 1993, \aj ~105, 1596

\bibitem[Chandrasekhar 1943]{C43}
Chandrasekhar, S. 1943, \apj 97, 255

\bibitem[Corcoran et al. 1994]{C94}
Corcoran, M.F., O'Neel, B., Perry, K., Smale, W., White, N., \& Petre, R. 1994
Rosat: The Images Volume 2 (HEASARC)

\bibitem[de Vaucouleurs et al. 1991]{RC3} 
de Vaucouleurs, G., de Vaucouleurs, A., Corwin, H., Buta, R.,
Paturel, G., \& Fouque, P. 1991 Third Reference Catalogue of Bright Galaxies
(Springer, New York)

\bibitem[den Hartog \& Katgert 1996]{HK96}
den Hartog, R. and Katgert, P. 1996, \mnras ~279, 349

\bibitem[Dressler \& Schectman 1988]{DS88}
Dressler, A. \& Schectman, S. 1988 \aj ~95, 985

\bibitem[Fadda et al. 1996]{F96}
Fadda, D., Girardi, M., Giuricin, G., Mardirossian, F., \& Mezzetti, M. 1996
\apj ~473, 670

\bibitem[Fabricant et al. 1993]{FAB93} 
Fabricant, D., Kurtz M.,
Geller M., Zabludoff A., Mack P., Wegner G. 1993 \aj ~105, 788

\bibitem[Fitchett M.J. 1988]{F88}
Fitchett, M.J. 1988 \mnras ~230, 169

\bibitem[Gebhardt \& Beers 1991]{GB91} 
Gebhardt, K. \& Beers, T.C. 1991 \apj ~383, 72

\bibitem[Gregorini et al. 1994]{G94}
Gregorini, L., de Ruiter, H.R., Parma, P., Sadler, E.M., Vettolani, G. \&
Ekers, R.D. 1994 \aaps ~106, 1

\bibitem[Heisler et al. 1985]{HTB85}
Heisler, J., Tremaine, S. \& Bahcall, J.N. 1985, \apj ~298, 8

\bibitem[Huchra et al. 1983]{HDTL} 
Huchra, J., Davis, M., Latham, D., \& Tonry, J. 1983 \apjsupp ~52, 89

\bibitem[Jacoby et al. 1984]{JHC84} 
Jacoby, G.H., Hunter, D.A. \& Christian, C.A. 1984 \apjsupp ~56, 257

\bibitem[Jones 1996]{J96} 
Jones, C. private communication

\bibitem[Jones \& Forman 1984]{JF84} 
Jones, C. \& Forman, W. 1984 \apj ~276, 38

\bibitem[Kaiser 1987]{K87}
Kaiser, N. 1987 \mnras, ~227, 1

\bibitem[Kinman \& Hintzen 1981]{KH} 
Kinman, T. \&  Hintzen, P. 1981 \pasp, ~93, 405

\bibitem[Kurtz et al. 1991]{K91} 
M.J. Kurtz, D.J. Mink, W.F. Wyatt, D.G. Fabricant, G. Torres, G.A. Kriss,
and J.L. Tonry 1991 in Astronomical Data Analysis Software and Systems I,
ASP Conf. Ser., Vol. 25, eds. D.M. Worrall, C. Biemesderfer, and J.
Barnes, p. 432-438.

\bibitem[Malumuth et al. 1992]{M92}
Malumuth, E., Kriss, G.A., Dixon, W.V.D., Ferguson, H.C. \& Ritchie,
C. 1992 \aj ~104, 495

\bibitem[Melnick \& Quintana 1981]{MQ81} 
Melnick J. \& Quintana, H.  1981 \aj ~86, 1567

\bibitem[Merrifield \& Kent 1989]{MK} 
Merrifield, M. R., \& Kent S. M. 1989 \aj ~89, 351

\bibitem[Owen et al. 1993]{OWG93}
Owen, F.N., White, R.A., \& Ge, J. 1993 \apjs ~87, 135

\bibitem[Pickles 1985]{PIC85} 
Pickles, A. J. 1985 \apj ~296, 340

\bibitem[Quintana \& Lawrie 1982]{QL82} 
Quintana, H. \& Lawrie 1982 \aj ~87, 1

\bibitem[Quintana et al. 1996]{QRW96} 
Quintana, H., Ramirez, A. \& Way, M.J. 1996 \aj ~112, 36

\bibitem[Sandage 1978]{S78} 
Sandage 1978 \aj ~83, 904

\bibitem[Shectman 1985]{SH85} 
Shectman S., Carnegie Institution of Washington Year Book 1989, p.25-32

\bibitem[Slee et al. 1989]{S89}
Slee, O.B., Perley, R.A. \& Siegman, B.C. 1989 Aust J. Phys. ~42, 633

\bibitem[Tody 1993]{T93}
Tody, D. 1993, "IRAF in the Nineties" in Astronomical Data Analysis Software
and Systems II, A.S.P. Conference Ser., Vol 52, eds. R.J. Hanisch,
R.J.V. Brissenden, and J. Barnes, 173. 

\bibitem[Tonry \& Davis 1979]{TD79} 
Tonry, J. \& Davis, M. 1979 \aj ~84, 1511 

\bibitem[West \& Bothun 1990]{WB90}
West, M.J. \& Bothun, G.D. 1990, \apj ~350, 36

\bibitem[Yahil \& Vidal 1977]{YV77}
Yahil, A., \& Vidal, N.V. 1977, \apj ~214, 347

\bibitem[Zabludoff et al. 1993]{ZAB93} 
Zabludoff, A., Geller, M., Huchra, J. \&  Vogeley, M. 1993, \aj ~106, 1273

\bibitem[Zhao et al. 1989]{ZBO89} 
Zhao, J., Burns, J.O., \& Owen, F.N. 1989, \aj ~98, 64

\end{thebibliography}
\end{document}


\begin{deluxetable}{cccrrrllrrr};
\tablenum{1}
\tablecaption{A119 - Velocities}
\tablehead{
\colhead{Ident.}           & \colhead{$\alpha$ (1950)} & 
\colhead{ $\delta$ (1950)} & \colhead{v$_{\odot}$}     &
\colhead{err}              & \colhead{R}               &
\colhead{ }                & \colhead{Ref}             & 
\colhead{Id.Ref}           & \colhead{ $\overline{\upsilon}_{\odot}$} &
\colhead{err}}
\startdata 
 25007 &  0 50 25.9 & -02 03 29.6 & 13586 &  66 &  4.5 &   & fibr &      &       &     \nl
 25003 &  0 50 39.1 & -01 04 16.2 & 40924 &  76 &  3.9 &   & fibr &      &       &     \nl
 25005 &  0 50 46.8 & -01 34 08.5 & 13148 &  45 &  6.2 &   & fibr &      &       &     \nl
 25012 &  0 50 55.9 & -00 56 53.8 & 13089 &  70 &  3.5 &   & fibr &      &       &     \nl
 25024 &  0 51 08.1 & -02 12 05.0 & 13782 &  29 & 10.1 &   & fibr &      &       &     \nl
 25020 &  0 51 09.2 & -01 21 22.2 & 13027 &  42 &  6.1 &   & fibr &      &       &     \nl
 25015 &  0 51 12.7 & -01 03 04.5 & 41345 &  57 &  4.9 &   & fibr &      &       &     \nl
 25017 &  0 51 13.4 & -01 09 14.7 & 14173 &  38 &  7.9 &   & fibr &      &       &     \nl
 25018 &  0 51 14.4 & -01 11 41.9 & 14082 &  89 &  2.4 &   & fibr &      &       &     \nl
 25027 &  0 51 18.4 & -02 11 16.3 & 20624 &  89 &  3.9 &   & fibr &      &       &     \nl
 25031 &  0 51 25.7 & -01 32 21.7 & 16058 &  50 &  6.1 &   & fibr &      &       &     \nl
 25030 &  0 51 25.7 & -01 43 26.0 & 13083 &  48 &  6.4 &   & fibr &      &       &     \nl
 25040 &  0 51 37.5 & -02 13 25.1 & 13480 &  39 &  8.7 &   & fibr &      &       &     \nl
 25038 &  0 51 38.8 & -01 41 57.8 & 13060 &  32 &  8.8 &   & fibr &      &       &     \nl
 25035 &  0 51 41.3 & -00 55 33.0 & 43573 & 130 &  2.9 &   & fibr &      &       &     \nl
 25045 &  0 51 42.2 & -02 04 04.7 & 13813 &  51 &  4.3 &   & fibr &      &       &     \nl
 25047 &  0 51 42.6 & -01 30 36.9 & 13506 &  31 &  8.7 &   & fibr &      &       &     \nl
 25049 &  0 51 52.7 & -01 11 56.5 & 12742 &  39 &  7.0 &   & fibr &      &       &     \nl
 25057 &  0 52 05.5 & -01 50 42.0 & 12666 &  42 &  6.0 &   & fibr &      &       &     \nl
 25058 &  0 52 07.7 & -01 50 44.3 & 12464 &  37 &  7.5 &   & fibr &      & 12481 &  30 \nl
       &            &             & 12534 &  44 &      &   & FAB93&   8  &       &     \nl
 25065 &  0 52 08.6 & -01 23 39.6 & 15944 &  53 &  5.1 &   & fibr &      &       &     \nl
 25062 &  0 52 08.9 & -01 45 35.9 & 11910 &  41 &  6.9 &   & fibr &      & 11875 &  32 \nl
       &            &             & 11873 &  38 &      &   & FAB93&  20  &       &     \nl
 25064 &  0 52 16.1 & -01 39 33.8 & 13989 &  31 &  9.5 &   & fibr &      & 13990 &  30 \nl
       &            &             & 14021 &  55 &      &   & FAB93&  25  &       &     \nl
 25063 &  0 52 17.4 & -01 45 00.4 & 13889 &  37 &  7.9 &   & fibr &      & 13905 &  30 \nl
       &            &             & 13959 &  46 &      &   & FAB93&  19  &       &     \nl
 25061 &  0 52 20.2 & -01 50 50.6 & 13995 &  31 &  8.5 &   & fibr &      & 14048 &  67 \nl
       &            &             & 14160 &  39 &      &   & FAB93&   7  &       &     \nl
 25076 &  0 52 22.5 & -01 43 24.6 & 29368 &  50 &  4.9 &   & fibr &      & 29381 &  45 \nl
       &            &             & 29462 & 100 &      &   & FAB93&  21  &       &     \nl
 25078 &  0 52 23.5 & -01 57 29.2 & 30284 &  98 &  5.1 &   & fibr &      &       &     \nl
 25077 &  0 52 23.9 & -01 51 19.6 & 32394 &  61 &  4.8 &   & fibr &      &       &     \nl
 25079 &  0 52 25.5 & -02 09 06.3 & 13681 &  45 &  6.8 &   & fibr &      &       &     \nl
 25075 &  0 52 30.2 & -01 35 17.1 & 13883 &  92 &      &   & FAB93&  56  &       &     \nl
 25073 &  0 52 31.7 & -01 31 23.6 & 13011 &  27 & 10.1 &   & fibr &      &       &     \nl
 25072 &  0 52 32.2 & -01 30 15.8 & 13351 &  56 &      &   & FAB93&  70  &       &     \nl
 25070 &  0 52 35.5 & -01 19 00.8 & 13777 &  26 & 10.9 &   & fibr &      & 13742 &  81 \nl
       &            &             & 13600 &  60 &      &   & MQ81 &  23  &       &     \nl
 25092 &  0 52 37.8 & -01 04 50.5 & 13663 &  36 &  7.6 &   & fibr &      &       &     \nl
 25088 &  0 52 40.3 & -01 20 46.8 & 14569 &  80 &  6.7 &   & fibr &      &       &     \nl
 25090 &  0 52 43.5 & -01 11 18.5 & 13242 &  32 &  9.1 &   & fibr &      &       &     \nl
 25086 &  0 52 44.6 & -01 33 24.8 & 13546 &  46 &    1 &   & fibr &      & 13532 &  33 \nl
       &            &             & 13546 &  46 &      &   & FAB93&  55  &       &     \nl
 25085 &  0 52 45.2 & -01 35 19.3 & 14166 &  84 &      &   & FAB93&  53  &       &     \nl
 25087 &  0 52 45.5 & -01 32 52.1 & 13427 &  30 & 11.7 &   & fibr &      & 13405 &  36 \nl
       &            &             & 13390 &  50 &      &   & MQ81 &   4  &       &     \nl
 25083 &  0 52 48.0 & -01 37 23.3 & 12490 &  42 &  7.0 &   & fibr &      &       &     \nl
 25099 &  0 52 49.6 & -01 28 53.8 & 12577 &  56 &  5.0 &   & fibr &      & 12594 &  42 \nl
       &            &             & 12645 &  65 &      &   & FAB93&  69  &       &     \nl
 25105 &  0 52 51.9 & -01 34 45.6 & 12472 &  27 & 12.1 &   & fibr &      & 12462 &  30 \nl
       &            &             & 12462 &  45 &      &   & FAB93&  51  &       &     \nl
 25094 &  0 52 52.3 & -01 00 45.5 & 13356 &  42 &  6.7 &   & fibr &      &       &     \nl
 25098 &  0 52 56.1 & -01 24 11.4 & 11887 &  26 & 11.6 &   & fibr &      & 11887 &  26 \nl
       &            &             & 19235 & 300 &      &   & MQ81 &  20  &       &     \nl
 25107 &  0 52 57.2 & -01 41 04.8 & 12873 &  34 &  9.7 &   & fibr &      & 12868 &  30 \nl
       &            &             & 12890 &  39 &      &   & FAB93&  24  &       &     \nl
 25093 &  0 52 58.2 & -00 48 02.3 & 13963 &  54 &  5.7 & e & fibr &      &       &     \nl
 25096 &  0 52 58.3 & -01 17 37.4 & 34667 &  47 &  5.8 &   & fibr &      &       &     \nl
 25100 &  0 52 58.9 & -01 28 53.7 & 13810 &  44 &  5.3 &   & fibr &      &       &     \nl
 25119 &  0 53 04.9 & -01 51 00.5 & 14692 &  49 &  6.8 &   & fibr &      &       &     \nl
 25122 &  0 53 06.6 & -01 45 34.6 & 12845 &  36 &  9.0 &   & fibr &      &       &     \nl
 25126 &  0 53 07.3 & -01 34 58.0 & 13017 &  35 &  7.8 &   & fibr &      & 13033 &  30 \nl
       &            &             & 13092 &  48 &      &   & FAB93&  50  &       &     \nl
 25137 &  0 53 09.7 & -01 18 15.2 & 12655 &  32 &  7.7 &   & fibr &      & 12658 &  30 \nl
       &            &             & 12694 &  48 &      &   & FAB93&  91  &       &     \nl
 25129 &  0 53 11.5 & -01 29 11.1 & 12529 &  51 &  4.4 &   & fibr &      &       &     \nl
 25125 &  0 53 11.9 & -01 35 42.0 & 13770 &  31 &  9.5 &   & fibr &      & 13746 &  41 \nl
       &            &             & 13705 &  52 &      &   & FAB93&  49  &       &     \nl
 25111 &  0 53 13.2 & -02 07 21.3 & 13190 &  58 &  4.8 &   & fibr &      &       &     \nl
 25116 &  0 53 14.5 & -01 58 24.0 & 12857 &  45 &  5.7 &   & fibr &      & 12906 &  47 \nl
       &            &             & 12980 &  44 &      &   & FAB93&   1  &       &     \nl
 25134 &  0 53 16.2 & -01 21 07.5 & 13411 &  32 &  8.7 &   & fibr &      &       &     \nl
 25148 &  0 53 17.7 & -01 26 06.0 & 14385 &  47 &  6.2 &   & fibr &      & 14330 &  48 \nl
       &            &             & 14316 &  41 &      &   & FAB93&  76  &       &     \nl
 25159 &  0 53 18.9 & -01 46 20.1 & 13104 &  33 &  9.3 &   & fibr &      & 13118 &  30 \nl
       &            &             & 13172 &  44 &      &   & FAB93&  18  &       &     \nl
 25144 &  0 53 20.1 & -01 10 16.6 & 13247 &  31 &  9.6 &   & fibr &      & 13282 &  90 \nl
       &            &             & 13560 &  80 &      &   & MQ81 &  17  &       &     \nl
 25146 &  0 53 20.8 & -01 11 35.0 & 13294 &  39 &  7.6 &   & fibr &      & 13268 &  33 \nl
       &            &             & 13270 &  50 &      &   & MQ81 &   9  &       &     \nl
 25145 &  0 53 21.1 & -01 11 25.7 & 13633 &  43 &  7.8 &   & fibr &      & 13676 &  70 \nl
       &            &             & 13835 &  70 &      &   & MQ81 &   8  &       &     \nl
 25151 &  0 53 22.3 & -01 31 03.0 & 12179 &  40 &  8.5 &   & fibr &      & 12189 &  30 \nl
       &            &             & 12190 &  33 & 10.6 &   & fibr &      &       &     \nl
       &            &             & 12320 & 130 &      &   & MQ81 &   5  &       &     \nl
 25154 &  0 53 25.7 & -01 34 14.6 & 14697 &  39 &  9.1 &   & fibr &      & 14692 &  30 \nl
       &            &             & 14712 &  48 &      &   & FAB93&  47  &       &     \nl
 25163 &  0 53 28.6 & -01 49 12.7 & 12705 &  40 &  7.1 &   & fibr &      & 12717 &  37 \nl
       &            &             & 12827 & 106 &      &   & FAB93&  15  &       &     \nl
 25157 &  0 53 29.3 & -01 36 16.9 & 12795 &  30 & 11.2 &   & fibr &      &       &     \nl
 99999 &  0 53 30.0 & -01 35 00.0 & 12851 &  73 &      &   & ZAB93&      &       &     \nl
 25143 &  0 53 29.8 & -01 05 53.8 & 13408 &  29 & 10.3 &   & fibr &      &       &     \nl
 25165 &  0 53 30.0 & -02 12 06.3 &  4048 &  25 &  5.0 & e & fibr &      &       &     \nl
 25201 &  0 53 31.8 & -00 43 30.8 & 12895 &  53 &  5.0 &   & fibr &      &       &     \nl
 25191 &  0 53 32.5 & -01 19 38.9 & 12949 &  41 &  6.5 &   & fibr &      & 12935 &  35 \nl
       &            &             & 12928 &  64 &      &   & FAB93&  87  &       &     \nl
 25170 &  0 53 33.7 & -01 36 49.2 & 13844 &  29 & 10.3 &   & fibr &      & 13836 &  30 \nl
       &            &             & 13836 &  54 &      &   & FAB93&  36  &       &     \nl
 25175 &  0 53 36.9 & -01 32 16.6 & 13690 &  56 &      &   & FAB93&  67  &       &     \nl
 25187 &  0 53 37.0 & -01 24 37.7 & 14517 &  29 &  9.4 &   & fibr &      & 14516 &  30 \nl
       &            &             & 14540 &  50 &      &   & FAB93&  82  &       &     \nl
 25190 &  0 53 37.3 & -01 23 13.7 & 12497 &  36 &  8.7 &   & fibr &      & 12501 &  33 \nl
       &            &             & 12565 &  80 &      &   & MQ81 &  11  &       &     \nl
 25166 &  0 53 38.0 & -01 48 06.0 & 14741 &  37 &  8.1 &   & fibr &      & 14736 &  30 \nl
       &            &             & 14758 &  42 &      &   & FAB93&  14  &       &     \nl
 25199 &  0 53 38.9 & -00 50 44.5 & 13230 &  70 &  3.5 &   & fibr &      &       &     \nl
 25186 &  0 53 40.9 & -01 24 52.2 & 13491 &  30 &  9.0 &   & fibr &      & 13476 &  30 \nl
       &            &             & 13452 &  56 &      &   & FAB93&  81  &       &     \nl
 25180 &  0 53 42.7 & -01 31 32.3 & 13326 &  35 & 10.1 &   & fibr &      & 13282 &  38 \nl
       &            &             & 13320 & 100 &      &   & MQ81 &   1  &       &     \nl
       &            &             & 13246 &  45 &      &   & RC3  &      &       &     \nl
 25220 &  0 53 42.9 & -01 35 01.9 & 12693 &  47 &    5 & e & fibr &      & 12711 &  30 \nl
       &            &             & 12752 &  40 &      &   & FAB93&  45  &       &     \nl
 25229 &  0 53 43.1 & -01 48 48.6 & 13109 &  46 &  7.6 &   & fibr &      & 13125 &  31 \nl
       &            &             & 13166 &  41 &      &   & FAB93&  13  &       &     \nl
 25207 &  0 53 43.6 & -01 13 24.6 & 42012 &  48 &  7.0 &   & fibr &      &       &     \nl
 25179 &  0 53 44.4 & -01 31 50.1 & 11712 &  41 &      &   & FAB93&  64  &       &     \nl
 25178 &  0 53 44.5 & -01 31 56.5 & 11554 &  86 &      &   & FAB93&  63  &       &     \nl
 25181 &  0 53 44.8 & -01 30 43.3 & 13131 &  36 &  9.4 &   & fibr &      & 13129 &  30 \nl
       &            &             & 13170 &  50 &      &   & MQ81 &  15  &       &     \nl
 25188 &  0 53 45.0 & -01 24 17.2 & 11061 &  48 &      &   & FAB93&  80  &       &     \nl
 25213 &  0 53 47.6 & -01 26 48.1 & 11730 &  34 &  8.8 &   & fibr &      & 11717 &  60 \nl
       &            &             & 11480 & 160 &      &   & MQ81 &  12  &       &     \nl
 25227 &  0 53 48.5 & -01 43 11.0 & 13933 &  44 &  5.6 &   & fibr &      &       &     \nl
 25214 &  0 53 49.5 & -01 28 47.8 & 14945 &  29 & 11.3 & e & fibr &      & 14873 & 175 \nl
       &            &             & 14705 &  50 &      &   & MQ81 &   3  &       &     \nl
 25208 &  0 53 50.1 & -01 15 25.4 & 13179 &  43 & 11.4 &   & fibr &      & 13185 &  39 \nl
       &            &             & 13255 &  90 &      &   & MQ81 &  16  &       &     \nl
 25212 &  0 53 51.2 & -01 24 47.5 & 12886 &  31 &  9.8 & e & fibr &      & 12839 &  77 \nl
       &            &             & 12760 &  50 &      &   & MQ81 &  13  &       &     \nl
 25217 &  0 53 52.2 & -01 31 57.1 & 11434 &  31 & 10.4 &   & fibr &      & 11456 &  30 \nl
       &            &             & 11575 & 250 &      &   & MQ81 &   2  &       &     \nl
       &            &             & 11541 &  48 &      &   & RC3  &      &       &     \nl
 25233 &  0 53 52.2 & -02 05 02.7 & 13866 &  43 &  6.0 &   & fibr &      &       &     \nl
 25218 &  0 53 52.5 & -01 32 42.6 & 13041 &  44 &  7.7 &   & fibr &      & 12995 &  80 \nl
       &            &             & 12765 &  80 &      &   & MQ81 &  14  &       &     \nl
       &            &             & 13069 &  49 &      &   & FAB93&  44  &       &     \nl
 25230 &  0 53 53.2 & -01 53 46.5 & 13706 &  37 &  8.4 &   & fibr &      & 13731 &  41 \nl
       &            &             & 13826 &  61 &      &   & FAB93&   5  &       &     \nl
 25206 &  0 53 53.9 & -01 03 45.0 & 12948 &  46 &  4.9 &   & fibr &      &       &     \nl
 25225 &  0 53 54.6 & -01 42 19.9 & 13458 &  32 &  9.3 &   & fibr &      & 13467 &  32 \nl
       &            &             & 13607 & 111 &      &   & FAB93&  23  &       &     \nl
 25216 &  0 53 55.8 & -01 29 49.1 & 11404 & 100 &      &   & FAB93&  59  &       &     \nl
 25241 &  0 53 57.1 & -01 48 15.1 & 14461 &  29 &  9.8 &   & fibr &      &       &     \nl
 25251 &  0 53 57.3 & -01 26 35.2 & 12657 &  29 & 10.0 &   & fibr &      & 12655 &  30 \nl
       &            &             & 12678 &  41 &      &   & FAB93&  74  &       &     \nl
 25239 &  0 53 58.1 & -01 53 12.0 & 13586 &  71 &      &   & FAB93&   4  &       &     \nl
 25245 &  0 53 58.2 & -01 37 54.9 & 12766 &  35 &  7.4 &   & fibr &      & 12758 &  32 \nl
       &            &             & 12750 &  76 &      &   & FAB93&  33  &       &     \nl
 25237 &  0 53 59.3 & -02  2 32.5 &  3970 & 100 &      &   & MQ81 &  21  &  4006 &  30 \nl
       &            &             &  4010 &  22 &      &   & HDTL &      &       &     \nl
 25250 &  0 54 02.3 & -01 32 08.7 & 13248 &  57 &  5.5 &   & fibr &      &       &     \nl
 25257 &  0 54 02.4 & -00 48 04.2 & 13258 &  61 &  4.1 &   & fibr &      &       &     \nl
 25240 &  0 54 03.9 & -01 48 36.8 & 13127 &  39 &  6.7 &   & fibr &      & 13142 &  36 \nl
       &            &             & 13247 &  88 &      &   & FAB93&  11  &       &     \nl
 25253 &  0 54 05.0 & -01 23 46.3 & 13535 &  33 &  7.9 &   & fibr &      &       &     \nl
 25244 &  0 54 05.2 & -01 39 38.2 & 13187 &  30 & 10.1 &   & fibr &      & 13198 &  42 \nl
       &            &             & 13386 & 114 &      &   & FAB93&  31  &       &     \nl
 25246 &  0 54 05.3 & -01 37 19.1 & 13383 &  36 &  8.8 &   & fibr &      & 13380 &  30 \nl
       &            &             & 13399 &  57 &      &   & FAB93&  32  &       &     \nl
 25248 &  0 54 05.7 & -01 33 55.4 & 15211 &  56 &  4.6 & e & fibr &      & 15079 & 150 \nl
       &            &             & 15022 &  45 &      &   & FAB93&  43  &       &     \nl
 25242 &  0 54 06.4 & -01 44 42.3 & 12908 &  28 & 10.0 &   & fibr &      &       &     \nl
 25254 &  0 54 07.5 & -01 18 16.5 & 13313 &  49 &  6.6 &   & fibr &      & 13270 &  38 \nl
       &            &             & 13265 &  43 &      &   & FAB93&  86  &       &     \nl
 25263 &  0 54 11.1 & -01 08 41.5 & 14064 &  34 &  8.8 &   & fibr &      & 14057 &  30 \nl
       &            &             & 14068 &  55 &      &   & FAB93& 108  &       &     \nl
 25281 &  0 54 12.8 & -01 48 51.5 & 12034 &  35 &  9.0 &   & fibr &      &       &     \nl
 25275 &  0 54 13.0 & -01 33 03.5 & 12610 &  30 &      &   & FAB93&  42  &       &     \nl
 25282 &  0 54 13.3 & -01 52 00.7 & 13456 &  41 &  6.2 &   & fibr &      & 13504 &  71 \nl
       &            &             & 13638 &  61 &      &   & FAB93&   3  &       &     \nl
 25262 &  0 54 13.6 & -01 08 53.1 & 13059 &  40 &  8.0 &   & fibr &      &       &     \nl
 25285 &  0 54 13.9 & -02 03 22.2 & 12391 &  65 &  3.8 & e & fibr &      &       &     \nl
 25258 &  0 54 15.9 & -00 53 44.3 & 13772 &  35 &  6.9 &   & fibr &      &       &     \nl
 25274 &  0 54 18.1 & -01 32 33.0 & 12245 &  30 & 11.1 &   & fibr &      & 12256 &  30 \nl
       &            &             & 12299 &  34 &      &   & FAB93&  41  &       &     \nl
 25271 &  0 54 19.3 & -01 25 44.8 & 13748 &  28 &  9.3 &   & fibr &      & 13724 &  30 \nl
       &            &             & 13689 &  74 &  3.0 &   & fibr &      &       &     \nl
       &            &             & 13699 &  47 &      &   & FAB93&  72  &       &     \nl
 25266 &  0 54 23.0 & -01 12 48.6 & 12791 &  42 &  7.1 &   & fibr &      &       &     \nl
 25280 &  0 54 23.4 & -01 43 23.1 & 13848 &  57 &  4.9 &   & fibr &      &       &     \nl
 25273 &  0 54 23.5 & -01 28 54.0 & 14964 &  76 &  2.9 & e & fibr &      & 15029 &  36 \nl
       &            &             & 15048 &  41 &      &   &  S78 &      &       &     \nl
 25289 &  0 54 23.7 & -01 39 31.9 & 13091 &  33 &  8.3 &   & fibr &      &       &     \nl
 25290 &  0 54 24.9 & -01 38 58.1 & 14806 &  42 &  6.4 &   & fibr &      &       &     \nl
 25294 &  0 54 25.2 & -01 31 37.1 & 13523 &  80 &      &   & FAB93&  57  &       &     \nl
 25301 &  0 54 26.8 & -01 05 42.8 & 14589 &  70 &  4.1 & e & fibr &      &       &     \nl
 25303 &  0 54 27.9 & -01 01 33.1 & 12984 &  50 &  5.8 &   & fibr &      &       &     \nl
 25298 &  0 54 28.5 & -01 08 59.4 & 13913 &  27 & 10.5 &   & fibr &      & 13911 &  30 \nl
       &            &             & 13935 &  39 &      &   & FAB93& 106  &       &     \nl
 99999 &  0 54 28.7 & -01 08 45.0 & 13535 &     &      &   & RC3  & 3405 &       &     \nl
 25299 &  0 54 29.0 & -01 08 42.4 & 13060 &  91 &  3.2 &   & fibr &      &       &     \nl
 25296 &  0 54 31.5 & -01 11 20.5 & 13451 &  53 &  5.5 &   & fibr &      & 13377 &  52 \nl
       &            &             & 13369 &  37 &      &   & FAB93&  97  &       &     \nl
 25288 &  0 54 33.6 & -01 40 03.0 & 14565 &  37 &  8.2 &   & fibr &      &       &     \nl
 25291 &  0 54 36.3 & -01 39 00.9 & 12010 &  31 & 10.7 &   & fibr &      &       &     \nl
 25319 &  0 54 37.2 & -01 17 19.2 & 12330 &  52 &  4.7 &   & fibr &      & 12393 & 108 \nl
       &            &             & 12625 &  90 &      &   & MQ81 &  18  &       &     \nl
 25292 &  0 54 37.8 & -01 39 00.1 & 14656 &  37 &  7.3 &   & fibr &      & 14651 &  32 \nl
       &            &             & 14665 &  62 &      &   & FAB93&  28  &       &     \nl
 25324 &  0 54 38.7 & -02 01 42.5 & 23761 &  68 &  3.3 &   & fibr &      &       &     \nl
 25313 &  0 54 40.3 & -00 55 01.9 & 12654 &  31 &  9.9 &   & fibr &      & 12666 &  76 \nl
       &            &             & 13200 & 200 &      &   & MQ81 &  19  &       &     \nl
 25318 &  0 54 40.4 & -01 16 54.7 & 11752 &  36 &  9.0 &   & fibr &      & 11752 &  36 \nl
       &            &             & 14765 & 230 &      &   & MQ81 &  22  &       &     \nl
 25322 &  0 54 41.2 & -01 33 20.5 & 12965 &  33 &  8.6 &   & fibr &      & 12963 &  30 \nl
       &            &             & 12988 &  41 &      &   & FAB93&  40  &       &     \nl
 25314 &  0 54 43.4 & -00 56 21.8 & 13518 &  56 &  6.1 &   & fibr &      & 13636 & 105 \nl
       &            &             & 13775 &  50 &      &   & MQ81 &   7  &       &     \nl
 25312 &  0 54 47.6 & -00 54 08.7 & 22821 &  89 &  2.8 &   & fibr &      &       &     \nl
 25315 &  0 54 51.3 & -01 00 36.0 & 18828 &  55 &  6.4 &   & fibr &      & 18793 &  38 \nl
       &            &             & 18788 &  53 &      &   & FAB93& 117  &       &     \nl
 25310 &  0 54 52.2 & -00 47 22.8 & 12669 &  47 &  6.8 &   & fibr &      &       &     \nl
 25329 &  0 54 52.9 & -01 36 48.3 & 13063 &  24 & 11.7 &   & fibr &      & 13076 &  63 \nl
       &            &             & 13411 & 117 &      &   & FAB93&  27  &       &     \nl
 25335 &  0 54 54.0 & -00 44 28.7 & 13156 &  53 &  7.2 &   & fibr &      &       &     \nl
 25331 &  0 54 58.4 & -01 11 51.6 & 13434 &  51 &      &   & FAB93&  96  &       &     \nl
 25328 &  0 55 01.5 & -01 39 39.6 & 13533 &  42 &  7.8 &   & fibr &      & 13491 &  34 \nl
       &            &             & 13467 &  71 &      &   & ZAB93&      &       &     \nl
       &            &             & 13465 &  60 &      &   & MQ81 &   6  &       &     \nl
 25350 &  0 55 03.0 & -02 08 02.1 & 23799 &  84 &  3.1 &   & fibr &      &       &     \nl
 99999 &  0 55 06.0 & -01 38 00.0 & 13427 &     &      &   & RC3  & 3444 &       &     \nl
 25344 &  0 55 06.4 & -01 09 44.2 & 13761 &  32 &  8.8 &   & fibr &      & 13745 &  39 \nl
       &            &             & 13678 &  77 &      &   & FAB93& 103  &       &     \nl
 25341 &  0 55 06.6 & -00 51 39.6 & 13680 &  66 &  3.7 &   & fibr &      &       &     \nl
 25347 &  0 55 08.3 & -01 34 55.2 & 14219 &  67 &  2.8 & e & fibr &      &       &     \nl
 25340 &  0 55 12.9 & -00 51 07.0 & 13893 &  68 &  5.2 &   & fibr &      &       &     \nl
 25345 &  0 55 15.3 & -01 16 25.3 & 12882 &  36 &  9.7 &   & fibr &      &       &     \nl
 25360 &  0 55 18.3 & -01 24 20.5 & 13091 &  53 &  5.2 & e & fibr &      &       &     \nl
 25361 &  0 55 20.2 & -01 05 03.7 & 12651 &  47 &  5.4 &   & fibr &      & 12620 &  40 \nl
       &            &             & 12598 &  59 &      &   & FAB93& 111  &       &     \nl
 25352 &  0 55 29.4 & -02 18 29.2 & 12814 &  41 &  6.8 &   & fibr &      &       &     \nl
 25362 &  0 55 31.6 & -00 57 32.3 & 12694 &  64 &  3.8 &   & fibr &      &       &     \nl
 25374 &  0 55 44.6 & -01 25 06.2 & 15647 &  38 & 10.2 &   & fibr &      & 15609 &  43 \nl
       &            &             & 15562 &  42 &  7.5 &   & fibr &      &       &     \nl
 25368 &  0 55 47.4 & -02 12 23.7 & 13018 &  37 &  8.0 &   & fibr &      &       &     \nl
 25372 &  0 55 51.3 & -01 39 51.0 & 15468 &  75 &  3.8 & e & fibr &      & 15473 &  70 \nl
       &            &             & 15510 & 200 &      &   &  KH  &      &       &     \nl
 25370 &  0 55 51.5 & -02 05 49.9 & 12661 &  42 &  7.8 &   & fibr &      & 12613 &  58 \nl
       &            &             & 12544 &  50 &  5.7 &   & fibr &      &       &     \nl
 25381 &  0 55 58.7 & -01 50 20.3 & 12801 &  74 &  3.4 &   & fibr &      &       &     \nl
 25380 &  0 56 03.6 & -01 05 28.0 & 22697 & 131 &    3 & e & fibr &      &       &     \nl
 25386 &  0 56 13.3 & -01 01 16.5 & 13333 &  66 &  5.1 &   & fibr &      &       &     \nl
 25385 &  0 56 14.1 & -01 21 59.5 & 13935 &  56 &  4.5 & e & fibr &      & 13939 &  40 \nl
       &            &             & 13942 &  56 &    6 & e & fibr &      &       &     \nl
 25389 &  0 56 24.9 & -01 06 18.0 & 13631 &  34 &  7.8 &   & fibr &      &       &     \nl
 25392 &  0 56 34.1 & -02 08 26.0 & 24554 &  47 &  5.9 &   & fibr &      &       &     \nl
 25393 &  0 56 37.4 & -02 10 07.5 & 15582 &  77 &  3.6 & e & fibr &      &       &     \nl
 25400 &  0 56 42.2 & -00 47 51.8 & 52499 &  95 &  3.3 &   & fibr &      &       &     \nl
 25398 &  0 56 45.2 & -01 15 25.8 & 13690 &  36 &  7.9 &   & fibr &      &       &     \nl
\tablerefs{
This paper: fibr = LCO fiber spectra; FAB93 = Fabricant et al. (1993);
HDTL = Huchra et al. (1983); MQ81 = Melnick \& Quintana (1981);
KH=Kinman \& Hintzen (1981); RC3 = de Vaucouleurs et al. (1991);
S78 = Sandage (1978); ZAB93 = Zabludoff et al. (1993)
}
\enddata
\end{deluxetable}


\begin{deluxetable}{cccrrrllrrr};
\tablecolumns{11}
\tablenum{2}
\tablecaption{A133 - Velocities}
\tablehead{
\colhead{Ident.}           & \colhead{$\alpha$ (1950)} & 
\colhead{ $\delta$ (1950)} & \colhead{v$_{\odot}$}     &
\colhead{err}              & \colhead{R}               &
\colhead{ }                & \colhead{Ref}             & 
\colhead{Id.Ref}           & \colhead{ $\overline{\upsilon}_{\odot}$} &
\colhead{err}}
\startdata 
 25022 &  0 57 10.1 & -21 39 48.7 & 16587 &  52 &  4.8 & e & fibr &      &       &     \nl
 25025 &  0 57 29.8 & -21 49 12.6 & 15952 &  49 &  5.7 &   & fibr &      &       &     \nl
 25026 &  0 57 31.5 & -21 47 54.6 & 17433 &  46 &  6.4 &   & fibr &      &       &     \nl
 25024 &  0 57 34.9 & -22 02 21.7 & 16400 &  50 &  5.7 &   & fibr &      &       &     \nl
 25027 &  0 57 35.0 & -21 45 27.5 & 16734 &  40 &  9.5 &   & fibr &      & 16753 &  30 \nl
       &            &             & 16805 &     &      &   & RC3  & 3582 &       &     \nl
 25035 &  0 57 39.3 & -21 49 00.0 & 16459 &  51 &  3.9 &   & fibr &      &       &     \nl
 25033 &  0 57 42.8 & -21 30 01.9 &  5635 &  46 &    7 & e & fibr &      &  5630 &  30 \nl
       &            &             &  5629 &  23 &    5 & e & fibr &      &       &     \nl
 25034 &  0 57 48.2 & -21 32 11.3 & 18297 &  78 &  3.1 & e & fibr &      &       &     \nl
 25038 &  0 57 58.0 & -22 31 31.2 & 55579 &  39 &  7.8 &   & fibr &      &       &     \nl
 25044 &  0 57 58.2 & -21 50 13.7 & 16404 &  48 &  6.4 &   & fibr &      &       &     \nl
 25042 &  0 57 58.5 & -21 59 06.1 & 16170 & 151 &    3 &   & fibr &      &       &     \nl
 25040 &  0 58 00.9 & -22 18 27.3 & 16136 &  37 &  7.8 &   & fibr &      &       &     \nl
 25039 &  0 58 01.1 & -22 28 59.2 & 17249 &  48 &  4.9 &   & fibr &      &       &     \nl
 25041 &  0 58 05.8 & -22 01 03.0 & 25300 &  40 &    3 & e & fibr &      &       &     \nl
 25047 &  0 58 12.1 & -21 48 40.7 & 34366 &  36 &  7.7 &   & fibr &      &       &     \nl
 25048 &  0 58 13.0 & -21 52 04.0 & 17298 &  29 &  9.9 &   & fibr &      & 17307 &  30 \nl
       &            &             & 17325 &  40 &  8.0 &   & fibr &      &       &     \nl
 25045 &  0 58 18.4 & -21 42 10.0 & 16505 &  38 &  7.0 &   & fibr &      &       &     \nl
 25050 &  0 58 20.5 & -21 59 35.0 & 12536 &  32 & 10.4 &   & fibr &      &       &     \nl
 25054 &  0 58 23.4 & -22 03 50.2 & 40375 &  92 &  2.9 & e & fibr &      &       &     \nl
 25056 &  0 58 27.9 & -21 54 36.4 & 34675 &  54 &  5.2 &   & fibr &      &       &     \nl
 25057 &  0 58 30.1 & -21 40 29.4 & 17245 &  39 &  7.2 &   & fibr &      &       &     \nl
 25053 &  0 58 34.0 & -22 05 54.9 & 17680 &  40 &  6.7 &   & fibr &      &       &     \nl
 25055 &  0 58 35.6 & -21 56 41.0 & 16926 &  44 &  6.6 &   & fibr &      &       &     \nl
 25058 &  0 58 41.3 & -21 38 13.8 & 16692 &  36 &  7.8 &   & fibr &      &       &     \nl
 25059 &  0 58 48.9 & -22 24 47.3 & 16593 &  37 &  7.5 &   & fibr &      &       &     \nl
 25062 &  0 58 50.9 & -22 28 33.7 & 17632 &  84 &  3.0 &   & fibr &      &       &     \nl
 25061 &  0 58 51.1 & -22 31 23.1 & 16942 &  29 & 10.0 &   & fibr &      &       &     \nl
 25068 &  0 58 53.3 & -22 11 56.6 & 17005 &  37 &  6.8 &   & fibr &      &       &     \nl
 25065 &  0 58 54.3 & -22 24 37.4 & 24804 &  29 &    4 & e & fibr &      &       &     \nl
 25069 &  0 58 56.2 & -21 44 20.0 & 16853 &  29 &  9.2 &   & fibr &      &       &     \nl
 25070 &  0 58 57.9 & -21 28 38.0 & 16472 &  38 &  7.8 &   & fibr &      &       &     \nl
 25064 &  0 58 58.1 & -22 26 30.4 & 17475 &  32 &  8.7 &   & fibr &      &       &     \nl
 25066 &  0 58 58.2 & -22 19 42.0 & 17435 &  41 &  6.6 &   & fibr &      &       &     \nl
 25067 &  0 58 58.3 & -22 14 33.8 & 15776 &  64 &    5 & e & fibr &      &       &     \nl
 25071 &  0 59 00.1 & -21 28 38.9 &  5500 &  97 &    3 & e & fibr &      &       &     \nl
 25084 &  0 59 07.2 & -22 35 06.0 & 17120 &  51 &  4.6 &   & fibr &      &       &     \nl
 25081 &  0 59 09.3 & -22 19 56.7 & 16020 &  36 & 11.2 &   & fibr &      &       &     \nl
 25082 &  0 59 11.6 & -22 22 49.5 & 16770 &  38 &  7.6 &   & fibr &      &       &     \nl
 25077 &  0 59 15.0 & -22 06 40.8 & 16322 &  27 &  9.5 &   & fibr &      &       &     \nl
 25083 &  0 59 15.2 & -22 28 31.8 & 17407 &  49 &  4.8 &   & fibr &      &       &     \nl
 25078 &  0 59 15.7 & -22 13 11.1 & 16327 & 113 &  3.0 &   & fibr &      &       &     \nl
 25074 &  0 59 17.6 & -21 22 50.3 & 16652 &  50 &  6.6 &   & fibr &      &       &     \nl
 25075 &  0 59 17.6 & -21 23 07.2 & 16625 &  42 &  7.0 &   & fibr &      &       &     \nl
 25079 &  0 59 20.6 & -22 16 29.6 & 16093 &  32 &  7.5 &   & fibr &      &       &     \nl
 25085 &  0 59 20.3 & -22 35 09.8 & 15603 &  31 &    5 & e & fibr &      & 15585 &  33 \nl
       &            &             & 15528 &  55 &    5 & e & fibr &      &       &     \nl
 25090 &  0 59 23.4 & -22 25 40.7 & 15299 &  36 &  6.8 &   & fibr &      &       &     \nl
 25086 &  0 59 28.6 & -22 32 13.2 & 15630 &  31 &    3 & e & fibr &      &       &     \nl
 25093 &  0 59 31.4 & -21 39 26.1 & 16313 &  60 &  5.0 &   & fibr &      &       &     \nl
 25095 &  0 59 31.9 & -21 21 53.2 & 30779 &  52 &  5.8 &   & fibr &      &       &     \nl
 25091 &  0 59 32.2 & -22 24 27.4 & 16675 &  39 &  7.7 &   & fibr &      &       &     \nl
 25088 &  0 59 34.3 & -22 30 58.6 & 17912 &  50 &  4.7 & e & fibr &      &       &     \nl
 99999 &  0 59 36.0 & -19 43 06.0 & 16891 &  33 &      &   & RC3  & 3695 &       &     \nl
 25112 &  0 59 39.2 & -22 22 25.8 & 16403 &  58 &  4.7 & e & fibr &      &       &     \nl
 25115 &  0 59 40.4 & -22 27 52.5 & 17492 &  27 & 11.3 &   & fibr &      &       &     \nl
 25111 &  0 59 40.6 & -22 20 49.1 & 15619 & 104 &  3.4 & e & fibr &      &       &     \nl
 25100 &  0 59 43.2 & -21 56 58.7 & 12479 &  43 &    6 & e & fibr &      &       &     \nl
 25113 &  0 59 44.2 & -22 26 16.6 & 16839 &  41 &  7.8 &   & fibr &      &       &     \nl
 25097 &  0 59 44.4 & -21 30 36.5 & 16797 &  52 &  6.3 &   & fibr &      &       &     \nl
 25098 &  0 59 44.4 & -21 31 16.4 & 16482 &  77 &  5.0 &   & fibr &      &       &     \nl
 25101 &  0 59 45.5 & -22 07 19.8 & 16719 &  84 &  2.4 &   & fibr &      &       &     \nl
 25102 &  0 59 45.5 & -22 08 16.7 & 16315 &  39 &  7.2 & e & fibr &      &       &     \nl
 25105 &  0 59 47.8 & -22 12 27.4 & 18132 &  34 &  8.9 &   & fibr &      &       &     \nl
 25107 &  0 59 47.2 & -22 13 31.8 & 17216 & 149 &  2.7 &   & fibr &      &       &     \nl
 25108 &  0 59 48.2 & -22 15 12.3 & 15837 &  36 &  8.4 &   & fibr &      &       &     \nl
 25103 &  0 59 48.3 & -22 08 56.2 & 17216 &  44 &  7.4 &   & fibr &      &       &     \nl
 25110 &  0 59 48.4 & -22 16 47.6 & 15555 &  64 &  3.5 & e & fibr &      &       &     \nl
 99999 &  0 59 50.0 & -19 56 18.0 & 17927 &  30 &      &   & RC3  & 3705 &       &     \nl
 25106 &  0 59 50.1 & -22 12 07.8 & 16999 &  39 &  7.2 &   & fibr &      &       &     \nl
 25119 &  0 59 50.8 & -22 22 45.5 & 17056 &  47 &  7.1 &   & fibr &      &       &     \nl
 25124 &  0 59 51.3 & -22 06 09.4 & 17552 &  29 & 10.5 &   & fibr &      &       &     \nl
 25122 &  0 59 52.3 & -22 09 07.3 & 17373 &  46 &  6.9 &   & fibr &      &       &     \nl
 25118 &  0 59 53.7 & -22 30 36.0 & 16249 &  37 &  8.9 &   & fibr &      &       &     \nl
 25123 &  1 00 00.7 & -22 07 18.2 & 16263 &  27 & 11.3 &   & fibr &      &       &     \nl
 25120 &  1 00 01.7 & -22 15 04.8 & 16985 &  37 &  9.0 &   & fibr &      &       &     \nl
 25130 &  1 00 03.6 & -21 47 16.9 & 17125 &  33 & 10.0 &   & fibr &      &       &     \nl
 25133 &  1 00 03.6 & -22 05 15.0 & 16371 &  38 &  8.0 &   & fibr &      &       &     \nl
 25140 &  1 00 04.9 & -22 06 27.8 & 16390 &  56 &  6.0 &   & fibr &      &       &     \nl
 25142 &  1 00 05.4 & -22 07 33.1 & 12024 &  72 &  4.8 &   & fibr &      &       &     \nl
 25132 &  1 00 07.0 & -21 53 26.1 & 28154 &  36 &  7.8 & e & fibr &      &       &     \nl
 25144 &  1 00 08.4 & -22 08 44.7 & 16846 &  31 & 10.4 &   & fibr &      & 16856 &  30 \nl
       &            &             & 17053 &  80 &      &   & MK   &      &       &     \nl
 25148 &  1 00 10.7 & -22 10 42.0 & 15521 &  80 &      &   & MK   &      &       &     \nl
 25147 &  1 00 11.0 & -22 09 11.0 & 16885 &  80 &      &   & MK   &      &       &     \nl
 25143 &  1 00 11.1 & -22 08 22.8 & 15415 &  30 & 10.6 &   & fibr &      &       &     \nl
 25151 &  1 00 11.9 & -22 14 58.7 & 16896 &  46 &  7.3 &   & fibr &      &       &     \nl
 25136 &  1 00 12.4 & -22 06 30.7 & 16409 &  40 &  8.8 &   & fibr &      &       &     \nl
 25146 &  1 00 12.9 & -22 09 04.7 & 17635 &  31 &  7.7 &   & fibr &      & 17629 &  30 \nl
       &            &             & 17722 &  80 &      &   & MK   &      &       &     \nl
 25128 &  1 00 13.3 & -21 36 17.3 & 16907 &  47 &  7.6 &   & fibr &      & 16976 &  78 \nl
       &            &             & 17064 &  53 &  5.6 &   & fibr &      &       &     \nl
 25149 &  1 00 13.7 & -22 13 35.2 & 17481 &  36 &  9.3 &   & fibr &      &       &     \nl
 25134 &  1 00 14.7 & -22 05 44.9 & 16462 &  33 &  8.3 &   & fibr &      &       &     \nl
 99999 &  1 00 15.7 & -22 08 09.0 & 17551 &  80 &      &   & MK   &      &       &     \nl
 25145 &  1 00 15.2 & -22 09 01.2 & 17052 &  47 & 10.2 &   & fibr &      & 17051 &  30 \nl
       &            &             & 17089 &     &      &   & RC3  & 3727 &       &     \nl
       &            &             & 17160 & 100 &      &   & MK   &      &       &     \nl
 99999 &  1 00 18.0 & -22 38 00.0 & 35376 &     &      &   & RC3  & 3730 &       &     \nl
 25150 &  1 00 18.8 & -22 13 36.1 & 16953 &  25 & 12.0 &   & fibr &      &       &     \nl
 25157 &  1 00 19.3 & -22 11 26.4 & 18475 &  47 &  7.3 &   & fibr &      &       &     \nl
 25171 &  1 00 21.9 & -21 52 09.7 & 16566 &  33 &  9.2 &   & fibr &      &       &     \nl
 25167 &  1 00 26.8 & -22 00 49.9 & 18326 &  39 &  7.4 &   & fibr &      &       &     \nl
 25153 &  1 00 28.4 & -22 35 31.9 & 16488 &  38 &  7.8 &   & fibr &      &       &     \nl
 25170 &  1 00 28.9 & -21 55 29.3 & 16206 &  42 &  6.3 &   & fibr &      &       &     \nl
 25155 &  1 00 29.9 & -22 16 49.0 & 16746 &  24 & 11.1 &   & fibr &      &       &     \nl
 25158 &  1 00 31.3 & -22 10 12.1 & 17224 &  46 &  7.4 &   & fibr &      &       &     \nl
 25190 &  1 00 33.1 & -22 27 02.7 & 15555 &  39 &  8.4 &   & fibr &      &       &     \nl
 25182 &  1 00 33.4 & -21 58 53.0 & 17195 &  43 &  6.6 &   & fibr &      &       &     \nl
 25168 &  1 00 34.3 & -21 56 48.2 & 17474 &  42 &  7.9 &   & fibr &      &       &     \nl
 25179 &  1 00 36.0 & -21 30 25.0 & 17860 &  99 &  3.2 & e & fibr &      &       &     \nl
 25169 &  1 00 36.2 & -21 55 48.1 & 16649 &  28 &  9.0 &   & fibr &      &       &     \nl
 25180 &  1 00 36.7 & -21 39 42.9 & 15813 &  54 &  3.2 & e & fibr &      &       &     \nl
 25181 &  1 00 37.1 & -21 46 47.4 & 17500 &  37 &  8.4 &   & fibr &      &       &     \nl
 25178 &  1 00 38.5 & -21 24 29.2 & 16022 &  58 &    4 & e & fibr &      &       &     \nl
 25184 &  1 00 45.1 & -22 13 30.7 & 15995 &  29 & 10.8 &   & fibr &      & 15985 &  30 \nl
       &            &             & 15961 &  45 &  5.6 &   & fibr &      &       &     \nl
 25189 &  1 00 45.9 & -22 21 13.1 & 18307 &  30 & 10.1 &   & fibr &      &       &     \nl
 25187 &  1 00 46.3 & -22 19 09.9 & 18062 &  31 &  8.8 &   & fibr &      &       &     \nl
 25188 &  1 00 46.5 & -22 19 59.2 & 18405 &  42 &  8.8 &   & fibr &      &       &     \nl
 25198 &  1 00 49.7 & -21 24 38.4 & 17716 &  62 &  5.1 &   & fibr &      &       &     \nl
 25196 &  1 00 50.7 & -22  0 51.7 & 15961 &  69 &  4.1 &   & fibr &      &       &     \nl
 25201 &  1 01 05.8 & -21 42 28.7 & 16172 &  45 &  5.3 &   & fibr &      &       &     \nl
 25200 &  1 01 06.9 & -21 26 59.4 & 25883 &  50 &  5.4 &   & fibr &      &       &     \nl
 25199 &  1 01 07.7 & -21 26 25.8 & 25970 &  65 &  5.3 &   & fibr &      &       &     \nl
 25203 &  1 01 09.7 & -22 24 19.2 & 17420 &  45 &  6.8 &   & fibr &      &       &     \nl
 25204 &  1 01 14.5 & -22 30 31.1 & 42414 &  63 &  4.2 &   & fibr &      &       &     \nl
 25210 &  1 01 18.4 & -21 32 16.9 & 17500 &  30 &  9.5 &   & fibr &      &       &     \nl
 25206 &  1 01 20.3 & -22 18 47.0 & 35765 &  45 &  7.7 &   & fibr &      &       &     \nl
 25208 &  1 01 28.1 & -22 10 58.8 & 36588 &  36 &  8.3 &   & fibr &      &       &     \nl
 25209 &  1 01 32.4 & -22 11 27.8 & 36907 &  49 &  5.6 &   & fibr &      &       &     \nl
 25212 &  1 01 33.3 & -21 38 27.1 & 17594 &  57 &  5.0 &   & fibr &      &       &     \nl
 25214 &  1 01 39.5 & -21 53 03.9 & 16228 &  82 &  3.5 & e & fibr &      &       &     \nl
 25217 &  1 01 53.3 & -22 20 51.3 & 18846 &  39 &  7.8 &   & fibr &      &       &     \nl
 25219 &  1 01 53.9 & -22 02 28.8 & 16865 &  32 &  8.5 &   & fibr &      &       &     \nl
 25215 &  1 01 54.1 & -22 30 08.4 & 18434 &  38 &  8.0 &   & fibr &      &       &     \nl
 25220 &  1 01 54.2 & -21 40 15.5 & 17241 &  55 &  6.1 &   & fibr &      &       &     \nl
 25216 &  1 01 54.8 & -22 29 20.9 & 18494 &  21 &    4 & e & fibr &      &       &     \nl
 25221 &  1 01 57.4 & -21 33 52.0 & 12394 &  81 &    4 & e & fibr &      &       &     \nl
 25224 &  1 01 60.0 & -21 48 30.3 & 16715 &  43 &  6.6 &   & fibr &      &       &     \nl
 25225 &  1 02 00.8 & -21 48 33.0 & 17518 & 118 &    3 & e & fibr &      &       &     \nl
 25227 &  1 02 02.9 & -22 13 39.9 & 37480 &  57 &  4.7 & e & fibr &      &       &     \nl
 25228 &  1 02 05.9 & -22 30 48.9 & 36933 &  44 &  6.1 &   & fibr &      &       &     \nl
 25223 &  1 02 06.1 & -21 39 24.2 & 17623 &  75 &  2.8 &   & fibr &      &       &     \nl
 25222 &  1 02 13.2 & -21 32 08.9 & 17473 &  55 &  6.1 &   & fibr &      &       &     \nl
 25238 &  1 02 16.8 & -21 35 50.2 & 17548 &  56 &  5.5 &   & fibr &      &       &     \nl
 25235 &  1 02 17.5 & -22 04 19.6 & 16732 &  39 &  7.4 &   & fibr &      &       &     \nl
 25229 &  1 02 19.2 & -22 32 13.4 & 16285 &  33 &  8.3 &   & fibr &      &       &     \nl
 25239 &  1 02 24.1 & -21 32 17.2 &  8010 &  18 &    5 & e & fibr &      &       &     \nl
 25240 &  1 02 25.4 & -21 30 26.6 & 25827 &  65 &  4.0 & e & fibr &      &       &     \nl
 25230 &  1 02 27.4 & -22 30 33.4 & 23611 &  80 &    4 & e & fibr &      &       &     \nl
 25245 &  1 02 32.7 & -22 16 21.1 & 37378 &  56 &  4.5 &   & fibr &      &       &     \nl
 25243 &  1 02 39.8 & -21 35 06.8 & 17486 &  44 &  6.4 &   & fibr &      &       &     \nl
 25244 &  1 02 42.1 & -22 13 09.0 & 17034 &  51 &  6.6 &   & fibr &      &       &     \nl
 25247 &  1 02 44.2 & -22 19 31.1 & 16845 &  31 &  9.5 & e & fibr &      & 16859 &  30 \nl
       &            &             & 16896 &  51 &  5.7 & e & fibr &      &       &     \nl
 25248 &  1 02 54.0 & -21 41 02.8 & 15279 &  74 &  3.7 & e & fibr &      &       &     \nl
 25246 &  1 02 55.2 & -22 33 03.3 & 36938 &  37 &    3 & e & fibr &      &       &     \nl
 25252 &  1 03 08.1 & -21 44 51.8 & 12112 &  24 &    4 & e & fibr &      & 12095 &  30 \nl
       &            &             & 12047 &  40 &    7 & e & fibr &      &       &     \nl
 25253 &  1 03 08.5 & -22 21 24.2 & 17206 &  28 & 11.4 &   & fibr &      &       &     \nl
 25254 &  1 03 18.3 & -21 47 43.0 & 31166 &  42 &    5 & e & fibr &      &       &     \nl
 25256 &  1 03 20.1 & -21 27 59.6 &  8456 &  38 &  7.5 &   & fibr &      &       &     \nl
\tablerefs{
This paper: fibr = LCO fiber spectra; MK = Merrifield \& Kent (1989);
RC3 = de Vaucouleurs et al. (1991)
}
\enddata
\end{deluxetable}